\documentclass[aps,showpacs,amsmath,amssymb,twocolumn,pra,superscriptaddress,notitlepage]{revtex4-2}

\usepackage{qcircuit}
\usepackage{amsmath,bm}
\usepackage[dvips]{graphicx}
\usepackage{amsmath,amssymb,amsthm,mathrsfs,amsfonts,dsfont}
\usepackage{subfigure, epsfig}
\usepackage{braket}
\usepackage{bm}
\usepackage{enumerate}
\usepackage{color}
\usepackage{graphicx}
\usepackage{algorithm}
\usepackage{algorithmic}
\usepackage{braket}
\usepackage{comment}
\usepackage{appendix}
\usepackage{here}
\usepackage{tabularx}

\newcommand{\tr}{\mathrm{Tr}}

\begin{document}


\title{Quantum error mitigation for rotation symmetric bosonic codes \\
with symmetry expansion}


\author{Suguru Endo}
\email{suguru.endou.uc@hco.ntt.co.jp}
\affiliation{NTT Computer and Data Science Laboratories, NTT Corporation, Musashino, 180-8585, Tokyo, Japan}

\author{Yasunari Suzuki}
\email{yasunari.suzuki.gz@hco.ntt.co.jp}
\affiliation{NTT Computer and Data Science Laboratories, NTT Corporation, Musashino, 180-8585, Tokyo, Japan}

\author{Kento Tsubouchi}
\affiliation{Department of Applied Physics, University of Tokyo,
7-3-1 Hongo, Bunkyo-ku, Tokyo 113-8656, Japan}

\author{Rui Asaoka}
\affiliation{NTT Computer and Data Science Laboratories, NTT Corporation, Musashino, 180-8585, Tokyo, Japan}

\author{Kaoru Yamamoto}
\affiliation{NTT Computer and Data Science Laboratories, NTT Corporation, Musashino, 180-8585, Tokyo, Japan}

\author{Yuichiro Matsuzaki}
\email{matsuzaki.yuichiro@aist.go.jp}
\affiliation{Research Center for Emerging Computing Technologies, National Institute of Advanced Industrial Science and Technology (AIST), Central2, 1-1-1 Umezono, Tsukuba, Ibaraki 305-8568, Japan}

\affiliation{
NEC-AIST Quantum Technology Cooperative Research Laboratory,
National Institute of Advanced Industrial Science and Technology (AIST), Tsukuba, Ibaraki 305-8568, Japan
}

\author{Yuuki Tokunaga}
\affiliation{NTT Computer and Data Science Laboratories, NTT Corporation, Musashino, 180-8585, Tokyo, Japan}


\begin{abstract}
\textcolor{black}{The rotation symmetric bosonic code (RSBC) is a unified framework of practical bosonic codes that have rotation symmetries, such as cat codes and binomial codes. While cat and binomial codes achieve the break-even point in which the coherence time of the encoded qubits exceeds that of unencoded qubits, the state preparation fidelity needs to be improved for practical quantum computing. Concerning this problem, we investigate the framework of symmetry expansion, a class of quantum error mitigation that virtually projects the state onto the noise-free symmetric subspace by exploiting the system's intrinsic symmetries and post-processing of measurement outcomes. Although symmetry expansion has been limited to error mitigation of quantum states immediately before measurement, we successfully generalize symmetry expansion for state preparation. Then, we consider two types of stabilization: photon number and phase stabilization. Photon number stabilization can be performed by leveraging the rotation operators. On the other hand, the phase errors can be suppressed via the extension towards phase direction by leveraging the recently proposed projective squeezing method. To implement our method, we use an ancilla qubit and a randomly generated controlled gate between the bosonic code states and the ancilla qubit. Our novel error mitigation method will significantly enhance computation accuracy in the near-term bosonic quantum computing. }

\end{abstract}

\maketitle

\section{introduction}
\label{section: introduction}

Quantum computers promise significant speedups for computation tasks such as Hamiltonian simulation, prime factoring, and machine learning~\cite{lloyd1996universal,shor1994algorithms,shor1999polynomial,harrow2009quantum,lloyd2014quantum}. However, detrimental effects of environmental noise devastates quantum advantages in such tasks. Quantum error correction with the redundancy of the code space
has been studied to correct computational errors to retrieve computation advantages~\cite{shor1995scheme,steane1996error,devitt2013quantum,lidar2013quantum,fowler2012surface}. 

For the realization of quantum error correction, discrete-variable-based code and continuous-variable-based code can be considered. \textcolor{black}{The former usually uses two-level systems as a basic component to encode a logical qubit. Although experimental progress is significant, e.g., distance-3 and -5 surface codes have been experimentally demonstrated by using superconducting qubit-systems~\cite{zhao2022realization,krinner2022realizing,acharya2022suppressing}, each logical qubit is encoded with tens of physical qubits, which require many connections of physical qubits and induce additional computation errors on the system due to engineering difficulties. As a result,
the break-even point has not been achieved, where the logical error rate is lower than the physical error rate of one physical qubit.}

In contrast, continuous-variable-based codes, which are also called bosonic codes, are hardware-efficient because only one bosonic mode in a resonator allows for the construction of a logical qubit from the infinite-dimensional Hilbert space~\cite{terhal2020towards,cai2021bosonic,joshi2021quantum,ma2021quantum,hu2019quantum}. Furthermore, for superconducting circuits, a microwave resonator has a longer coherence time than the qubit~\cite{reagor2016quantum}, so the error rate of the bosonic code tends to be lower. \textcolor{black}{Due to these advantages, the break-even point has already been observed for the bosonic cat code and the binomial code~\cite{ofek2016extending,ni2022beating}. In addition, for some bosonic codes, the noise channel is highly biased, and we can improve the threshold values by using error-correction protocols adapted for such noise~\cite{darmawan2021practical}.}

As a unified framework of the bosonic cat code~\cite{cochrane1999macroscopically} and binomial code~\cite{michael2016new}, the rotation symmetric bosonic code (RSBC) has been proposed by \textcite{grimsmo2020quantum}. RSBC code states are stabilized by rotation operations on the phase space represented by, e.g., Wigner functions~\cite{PhysRev.40.749,gerry2005introductory}, analogous to Gottesman-Kitaev-Preskill (GKP)  codes stabilized by displacement operations~\cite{gottesman2001encoding,grimsmo2021quantum}. The logical $Z$ operator can be described by the rotation operator $\hat{Z}_M=\mathrm{exp}(i \frac{\pi}{M} \hat{N})$, where $\hat{N}$ is the number operator and $M$ is the parameter to determines the degree of the RSBC. We note that $\hat{R}_M= \hat{Z}_M^2$ acts as a logical identity operator, i.e., the RSBC is invariant under the rotation with $2\pi/M$.

\textcolor{black}{Despite many advantages for error correction}, the state preparation noise 
cannot be ignored, which limits the practicality of the bosonic codes.
The state preparation fidelity of experiments for cat and binomial code is below 96 $\%$~\cite{ofek2016extending,hu2019quantum,vlastakis2013deterministically,ni2022beating}. State preparation of general RSBC states can be performed either with universal operations via dispersive interactions~\cite{krastanov2015universal} or consecutive projections via rotation operators with Hadamard test circuits~\cite{grimsmo2020quantum}. The former requires long pulse sequences. The latter needs high fidelity single-shot readout of the ancilla qubit; furthermore, the success probability of state preparation decreases as the rotation degree $M$ increases. Therefore, while preparing the encoded initial state, it is highly probable that the state is decohered due to noise, e.g., photon loss.

Meanwhile, low-overhead error suppression methods, so-called quantum error mitigation (QEM) techniques, have been intensively studied for leveraging noisy NISQ hardware for useful tasks~\cite{endo2021hybrid,cai2022quantum,qin2022overview,russo2022testing}.  There exist diverse QEM methods such as extrapolation~\cite{li2017efficient,temme2017error,Su2021errormitigationnear}, probabilistic error cancellation~\cite{temme2017error,endo2018practical,sun2021mitigating}, virtual distillation~\cite{koczor2021exponential,huggins2021virtual,huo2022dual}, and quantum subspace expansion~\cite{mcclean2017hybrid,yoshioka2022generalized}.  When syndrome measurements, decoding, and adaptive feedback controls are not available, one of QEM methods called symmetry expansion (SE) works as an alternative to error detection/correction techniques \textcolor{black}{by virtually projecting the noisy state onto the symmetric subspace with a random sampling of symmetry operators intrinsic to the system and post-processing of measurement outcomes~\cite{mcclean2020decoding,cai2021quantum}}.

\begin{figure*}[htp]
\begin{center}
    \includegraphics[width=2.0\columnwidth]{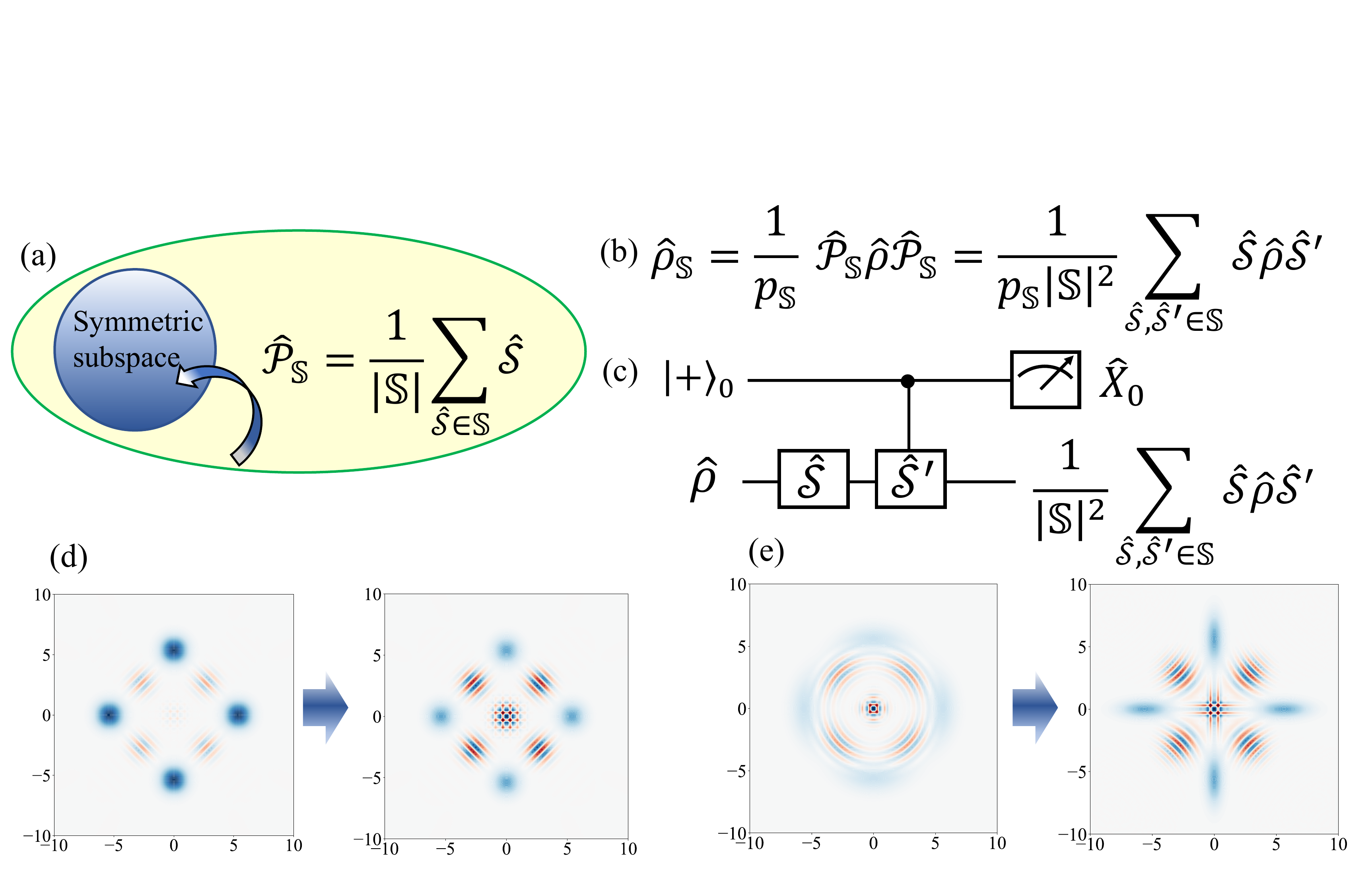}
    \caption{Schematic figures of symmetry expansion (SE) for rotation symmetric bosonic codes (RSBCs). (a) The concept of symmetric expansion (SE). We can project the noisy state onto the symmetric subspace with the projector $\hat{\mathcal{P}}_{\mathbb{S}}$, where $\mathbb{S}$ is the group of symmetry operations. The projector can be expanded as $\hat{\mathcal{P}}_{\mathbb{S}}=\frac{1}{|\hat{\mathcal{P}}_{\mathbb{S}}|}\sum_{\hat{\mathcal{S}}\in \mathbb{S}} \hat{\mathcal{S}}$ with symmetry operators $\hat{\mathcal{S}} \in \mathbb{S}$. In our case, we set $\hat{\mathcal{S}}$ to the rotation operators. (b) The error-mitigated state via SE, which is in the symmetric subspace. (c) The proposed quantum circuit for performing SE for the initial state preparation of RSBCs. We use an ancilla two-level system. The black/white circle indicates a control operation that operates when the ancilla qubit is $1/0$. Because we focus on the case where the symmetry operators are rotation operators, controlled-$\hat{\mathcal{S}}$ operations are dispersive interactions between the ancilla qubit and the system bosonic-code state. By randomly generating rotation operators and measuring a Pauli $X$ operator of the ancilla qubit, we can virtually obtain the effective density matrix projected onto the symmetric subspace. (d) The Wigner functions of a noisy logical zero state for the rotation order $M=2$ and the error-mitigated state. We simulate photon loss noise described by the Lindblad master equation $\frac{d \hat{\rho}}{dt}=\frac{\gamma}{2}(2\hat{a}\hat{\rho} \hat{a}^\dag -\hat{a}^\dag \hat{a} \hat{\rho} -\hat{\rho} \hat{a}^\dag \hat{a})$ for $\gamma t=0.1$. (e) The Wigner functions of a noisy state under phase error and the error-mitigated state for phase errors via projective squeezing. We simulate $\frac{d \hat{\rho}}{dt }=\frac{\gamma}{2}(2 \hat{N} \hat{\rho}  \hat{N}- \hat{N}^2\hat{\rho}-\hat{\rho} \hat{N}^2 )$ for $\gamma t=0.1$ with the increased squeezing level $\Delta r= 0.589$.  }
    \label{Fig:largefig1}
\end{center}
\end{figure*}

In this paper, we show how to apply SE for RSBCs. Although the conventional SE is applied just before the measurement for mitigating errors during computation, we also make SE applicable to the mitigation of state preparation noise with a constant depth by using a single ancilla qubit and controlled operations between the ancilla qubit and the resonator. The key point of our method is to exploit the recently proposed generalized quantum process used in the perturbative quantum simulation method, where we virtually add interactions to the system Hamiltonian~\cite{sun2022perturbative}. The schematic figures illustrating our proposal are displayed in Fig. \ref{Fig:largefig1}.  By using the same procedure, we can also virtually simulate RSBCs from easy-to-prepare states such as coherent states. We also showed that SE immediately before the measurement is useful when the information on symmetries is not available in the measurement. 

\textcolor{black}{We consider two types of stabilization: photon-number stabilization and phase stabilization. Photon-number-changing errors can be suppressed using rotation symmetries, while phase errors can be mitigated through photon-number translation symmetries. We demonstrate that stabilization via rotation symmetries can be achieved using our SE technique, effectively suppressing the impact of photon loss. For phase errors, although leveraging photon-number translation symmetries presents experimental challenges, the recently proposed projective squeezing method, which is a generalization of SE for translation symmetries, enables stretching the quantum state in the phase direction, thereby suppressing phase errors. We confirm numerically and analytically that our method significantly enhances computational accuracy. }

\section{Rotation symmetric bosonic code}
Here, we review the rotation symmetric bosonic code (RSBC), which is defined by the rotation symmetries in the phase space \cite{grimsmo2020quantum}. The logical states of the order-$M$ rotation symmetric code are defined as:
\begin{equation}
\begin{aligned}
\ket{0_{M,\Phi}}&={\mathcal{C}_0}^{-1/2} \sum_{k=0}^{2M-1} \mathrm{exp}\bigg(i \frac{k \pi }{ M} \hat{N}\bigg) \ket{\Phi} \\
\ket{1_{M,\Phi}}&={\mathcal{C}_1}^{-1/2} \sum_{k=0}^{2M-1} (-1)^k \mathrm{exp}\bigg(i \frac{k\pi}{M} \hat{N}\bigg) \ket{\Phi}, 
\end{aligned}
\label{Eq: rotation}
\end{equation}
where $\ket{\Phi}$ is a primitive state we need to choose properly, and $\mathcal{C}_0$ and $\mathcal{C}_1$ are normalization constants which asymptotically approaches $2M$ in the large photon limit, and $\hat{N}=\hat{a}^\dag \hat{a}$ is a number operator. Here, $\hat{a}$ and $\hat{a}^\dag$ denote annihilation and creation operators. These two states are stabilized by a rotation operator $\hat{R}_M=\mathrm{exp}(i \frac{2\pi}{M} \hat{N})$ with $\hat{Z}_M=\hat{R}_{2M}=\mathrm{exp}(i \frac{\pi}{M} \hat{N})$ operating as a logical $Z$ operator. 

Through the Kronecker comb relation $\frac{1}{M}\sum_{m=0}^{M-1}e^{i \frac{2\pi k m}{M}}=\sum_{l=0}^{\infty} \delta_{k, lM}~(k=0,1,2...)$, the projector onto the subspace spanned by Fock states $\{\ket{2nM + l} \}_n$  is constructed as 

\begin{equation}
\begin{aligned}
\hat{\mathcal{P}}_{2 M}^{(l)} &= \sum_{n=0}^\infty \ket{2nM +l}\bra{2nM+l} \\
&= \frac{1}{2 M} \sum_{k=0}^{2M-1}e^{\frac{-i\pi l k} {M} } \hat{Z}_M^k, 
\label{Eq: projectorrotation}
\end{aligned}
\end{equation}
where $l\in \{1,2,...,2M-1 \}$. Thus, the projectors onto $\ket{0_{M,\Phi}}$ and
$\ket{1_{M,\Phi}}$ correspond to $l=0$ and $l=M$. So, $\ket{0_{M,\Phi}}$ and $\ket{1_{M,\Phi}}$ has a Fock basis representation
\begin{equation}
\begin{aligned}
\ket{0_{M,\Phi}} &= \sum_{n=0}^{\infty} c_{2 n M}^{(\Phi)} \ket{2 n M} \\
\ket{1_{M,\Phi}} &= \sum_{n=0}^{\infty} c_{(2 n+1) M}^{(\Phi)} \ket{(2 n+1) M}. 
\end{aligned}
\label{Eq:fockbasis}
\end{equation}
Here, $c_i^{(\Phi)}$ is a probability amplitude dependent on the primitive state $\ket{\Phi}$. The projector on the code space can be written as
\begin{equation}
\begin{aligned}
\hat{\mathcal{P}}_{2M}^{(\mathrm{c})}&=\hat{\mathcal{P}}_{2 M}^{(0)}+\hat{\mathcal{P}}_{2 M}^{(M)} \\
&=\frac{1}{M}\sum_{k=0}^{M-1} \hat{R}_M^k.
\end{aligned}
\end{equation}

Note that, unlike the logical $Z$ operation, finding a simple form of the logical $X$ operation is not so straightforward. The dual states $\ket{\pm _{M,\Phi}}=\frac{1}{\sqrt{2}}(\ket{0_{M,\Phi}}\pm\ket{1_{M,\Phi}})$ are written in the Fock basis as:
\begin{equation}
\begin{aligned}
\ket{+_{M,\Phi}}&=\frac{1}{\sqrt{2}}\sum_{n=0}^{\infty} c_{nM}^{(\Phi)} \ket{nM} \\
\ket{-_{M,\Phi}}&=\frac{1}{\sqrt{2}}\sum_{n=0}^{\infty}(-1)^n c_{nM}^{(\Phi)} \ket{nM}.
\end{aligned}
\end{equation}

The important classes of the RSCBs are the cat code and the binomial code. For the cat code, the primitive state is the coherent state $\ket{\alpha}$, and the code states are the superposition of coherent states with different phases. The binomial code is defined as
\begin{equation}
\ket{0/1}_{\rm binom}=\frac{1}{\sqrt{2^L}} \sum_{m~ \mathrm{even/odd}}^{[0, L+1]} \sqrt{\binom{L+1}{m}} \ket{m M}, 
\end{equation}
where $M$ corresponds to the spacing in the Fock basis, and $L$ defines the truncation of the photon number. The binomial codes can exactly correct photon loss, photon gain, and dephasing errors up to a certain order. Note that the binomial code words $\ket{0/1}_{\rm binom}$ are stabilized by $\hat{\mathcal{P}}_{2 M}^{(0)}$ and $\hat{\mathcal{P}}_{2 M}^{(M)}$ due to the spacing structure in the Fock basis; thus the binomial code can be classified into the RSBCs. The primitive state of the binomial code is shown in Ref. \cite{grimsmo2020quantum}.

 With respect to the logical $Z$ measurement for RSBCs, Eq. (\ref{Eq:fockbasis}) indicates that a photon-number resolving measurement can distinguish the two states; more concretely, if the measurement outcome of photon number is $m M~(m=0, 1, 2...)$, the measured state turns out to be $\ket{0_{M,\Phi}}$ for even $m$, and vice versa. Thus, photon-number resolving measurement works as a logical $Z$ measurement on the code space. In the presence of noise \textcolor{black}{shifting the photon number}, by rounding the photon number to the nearest $m M$, we can suppress the effect of errors. Therefore, this code becomes increasingly robust to errors such as photon-loss or photon-gain errors.

For the logical $X$ measurement, in the large photon limit where $\mathcal{C}_0,  \mathcal{C}_1 \rightarrow 2M$, because the dual states can be rewritten from Eq. (\ref{Eq: rotation}) as
\begin{equation}
\begin{aligned}
\ket{+_{M,\Phi}} &\rightarrow \frac{1}{\sqrt{M}} \sum_{n=0}^{M-1} \mathrm{exp}\bigg(i\frac{2n\pi}{M}\hat{N}\bigg)\ket{\Phi} \\
\ket{-_{M,\Phi}} &\rightarrow \frac{1}{\sqrt{M}} \sum_{n=0}^{M-1} \mathrm{exp}\bigg(i\frac{(2n+1)\pi}{M} \hat{N}\bigg)\ket{\Phi},
\end{aligned}
\end{equation}
we can distinguish $\ket{+_{M,\Phi}}$ and $\ket{-_{M,\Phi}}$ with a phase measurement, e.g., heterodyne measurement~\cite{gardiner2004quantum}. The measurement operator corresponding to the heterodyne measurement is described by $\{\ket{\beta}\bra{\beta}\}_{\beta}$ for a coherent state $\ket{\beta}~(\beta \in \mathbb{C})$. When the measured angle $\theta=\mathrm{arg}(\beta)$ is closer to $2n\pi/M$, we can interpret the outcome as $+$, and vice versa. Note that while the robustness of the logical $Z$ measurement is increased as the rotation degree $M$ increases, the performance of the logical $X$ measurement becomes more vulnerable to phase noise. 

Even in the absence of phase noise, the failure probability of distinguishing the two logical states is finite with phase measurement due to the intrinsic phase uncertainty in quantum states. It has been shown that the phase uncertainty vanishes in the large photon number limit for the cat code ($\alpha \rightarrow \infty$) and the binomial code ($L\rightarrow \infty$). In this limit, the RSBC is called a number-phase code, in which the probability amplitudes are flattened, i.e., $|c_{nM}^{(\Phi)}|=|c_{(n+1)M}^{(\Phi)}|$ and the logical $X$ operator approaches a number-translation operator $\hat{X}_{\rm N}=\sum_{m=0}^\infty \ket{m}\bra{m+M}$. 

To perform universal operations, we need a universal gate set. The proposed universal gate set in number-phase codes is $\{\hat{H}, \hat{S}, \hat{T}, \hat{C} \}$ with $H$, $S$, $T$ denoting Hadamard, $S$, and $T$ gates, and $\hat{C}$ is a controlled rotation (CROT) gate between two bosonic modes defined as $\hat{C}_{M M'}=e^{i \frac{\pi}{M M'} \hat{N} \otimes \hat{N}}$. Here, $M$ and $M'$ denote the degree of rotation symmetry for two bosonic modes. Similarly to the Pauli stabilizer codes, a direct implementation of $T$ gates is not fault-tolerant, so we resort to the gate teleportation of the magic state $\ket{T_{M,\Phi}}= \frac{1}{\sqrt{2}}(\ket{0_{M,\Phi}}+e^{i\frac{\pi}{4}}\ket{1_{M,\Phi}})$  via a CROT gate together with the logical $X$ measurement for the gate operation. As well as the absence of a simple implementation of the logical $X$ operation, the Hadamard operation cannot be straightforwardly applied. Then we use the teleportation of the plus state $\ket{+_{M, \Phi}}$ for applying the Hadamard gate. The direct implementation of the $S$ gate is expressed as $\hat{S}=e^{i\frac{\pi}{2M^2} \hat{N}^2}$, which can be also performed via teleportation of $\ket{+i_{M,\Phi}}=\frac{1}{\sqrt{2}}(\ket{0_{M,\Phi}}+i\ket{1_{M,\Phi}})$. 

For the error correction of the number phase code, due to the complicated structure of the logical $X$ operator, we use a teleportation-based quantum error-correction scheme~\cite{knill2005scalable}: two fresh ancilla qubits encoded in RSBCs are added, and two CROT gates are applied, one of which is applied to the noisy state and one of the fresh qubits and the other is applied to the fresh ancilla qubits. Then, the logical $X$ measurements are performed to one of the ancilla qubits and the noisy state to teleport the state to one of the fresh ancilla qubits. \textcolor{black}{When the dual states are distinguishable in the presence of propagated noise in the logical $X$ measurement, we can obtain the error information for error correction in the teleported mode.} Refer to Ref. \cite{hillmann2022performance} for the detailed study of performance of teleportation-based error correction under noisy measurements.

\section{Symmetry expansion}
\label{Sec: SE}
The symmetry expansion (SE) method allows us to virtually project the noisy quantum state onto the symmetric subspace~\cite{mcclean2020decoding,cai2021quantum}.
Let us denote the \textcolor{black}{finite} group of symmetry operations by $\mathbb{S}$. Then, the state in the symmetry
subspace $\ket{\psi_{\mathbb S}}$ is stabilized as
\begin{equation}
\hat{\mathcal{S}} \ket{\psi_{\mathbb S}} = \ket{\psi_{\mathbb S}} ~\forall \hat{\mathcal{S}} \in \mathbb{S}.
\end{equation}
The projector onto the symmetric subspace reads:
\begin{equation}
\hat{\mathcal{P}}_{\mathbb{S}}=\frac{1}{|\mathbb{S}|} \sum_{\hat{\mathcal{S}} \in \mathbb{S}} \hat{\mathcal{S}}.
\end{equation}
Now, we suppose the noisy state $\hat{\rho}$ is projected via the projector: 
\begin{equation}
\hat{\rho}_{\mathbb{S}}= \frac{\hat{\mathcal{P}}_{\mathbb{S}} \hat{\rho} \hat{\mathcal{P}}_{\mathbb{S}}}{\mathrm{Tr}[\hat{\mathcal{P}}_{\mathbb{S}} \hat{\rho} ]}
\end{equation}

In SE, similarly to the
other error mitigation methods, we can mitigate errors in expectation values of observables via classical post-processing of measurement outcomes. Let the measured observable denote as $\hat{O}$, which we assume commutes with the projector $\hat{\mathcal{P}}_{\mathbb{S}}$. The error-mitigated expectation value results in
\begin{equation}
\begin{aligned}
\braket{\hat{O}}_{\mathbb{S}}&= \frac{1}{p_\mathbb{S} |\mathbb{S}|} \sum_{\hat{\mathcal{S}} \in \mathbb{S} } \mathrm{Tr}[\hat{\mathcal{S}} \hat{O} \hat{\rho}] \\
p_\mathbb{S}&= \frac{1}{ |\mathbb{S}|} \sum_{\hat{\mathcal{S}} \in \mathbb{S} } \mathrm{Tr}[\hat{\mathcal{S}} \hat{\rho}]. 
\end{aligned}
\end{equation}
We first compute $p_{\mathbb{S}}$ by uniformly sampling $\hat{\mathcal{S}} \in \mathbb{S}$, measuring it, and repeating this procedure to obtain the average of the outcomes. We can similarly compute $\frac{1}{|\mathbb{S}|} \sum_{\hat{\mathcal{S}} \in \mathbb{S} } \mathrm{Tr}[\hat{\mathcal{S}} \hat{O} \hat{\rho}]$ by sampling $\hat{\mathcal{S}}\hat{O}$ instead. Thus we can obtain $\braket{\hat{O}}_{\mathbb{S}}$. Although we focused on the case of uniformly sampling symmetries, we can also consider optimizing the weight of each symmetry. 

Let us now assume the symmetry is characterized by Pauli stabilizer generators $\{\hat{\mathcal{S}}_i \}_i$. In this case, the projector onto the symmetric subspace can be expressed as
\begin{equation}
\begin{aligned}
\hat{\mathcal{P}}_{\mathrm{Pauli}}&= \prod_i \bigg(\frac{I+\hat{\mathcal{S}}_i}{2}\bigg)\\
&=\frac{1}{2^{N_{\rm g}}} \sum_{\hat{\mathcal{S}} \in \mathbb{S}_{\rm stab}} \hat{\mathcal{S}}, 
\end{aligned}
\end{equation}
where $\mathbb{S}_{\rm stab}$ denotes the stabilizer group generated by $\{\hat{\mathcal{S}}_i \}_i$ with $N_{\mathrm{g}}$ being the number of stabilizer generators. Thus, SE can be performed by randomly sampling $\hat{\mathcal{S}} \in \mathbb{S}_{\rm stab}$ with the probability $1/{2^{N_{\rm g}}}$.

Now we discuss the sampling cost required in this method. Given random variables $x$ and $y$, the variance of the function $f(x,y)$ under the assumption that there is no correlation of $x$ and $y$ is
\begin{equation}
\mathrm{Var}[f]=\bigg|\frac{\partial f}{\partial x}\bigg|^2 \mathrm{Var}[x]+\bigg|\frac{\partial f}{\partial y}\bigg|^2 \mathrm{Var}[y].
\end{equation}
For $f(x,y)=x/y$, we have
\begin{equation}
\mathrm{Var}[f]= \frac{1}{\braket{y}^2} (\mathrm{Var}[x]+\braket{f}^2  \mathrm{Var}[y] ) \propto \braket{y}^{-2},
\label{Eq: propagation}
\end{equation}
where 
$\braket{x}$ and
$\mathrm{Var}[x]$ denotes the expectation value and variance of a random variable $x$. Thus, in the case of symmetry expansion, the variance of the error-mitigated expectation value $\braket{\hat{O}}_{\mathbb{S}}$ is proportional to $p_{\mathbb{S}}^{-2}$, and the number of measurements to achieve the required accuracy $\varepsilon$ scales as $O( (p_{\mathbb{S}}\varepsilon)^{-2} )$.

\section{Symmetry expansion for rotation symmetric bosonic codes}
In this section, we first introduce error mitigation for state preparation errors via SE using the recently introduced generalized process~\cite{sun2022perturbative}. \textcolor{black}{Next, we compare our method with the conventional state verification method.} We then discuss how SE immediately before measurement can be used for RSBCs. Finally, we show that RSBCs can be virtually generated by applying SE to a primitive state $\ket{\Phi}$.

\subsection{Symmetry expansion for state preparation}
Here, we introduce symmetry expansion (SE) for RSBCs for mitigation of state preparation errors, while SE just before measurement has only been discussed for Pauli and permutation symmetries~\cite{cai2021quantum,mcclean2020decoding}. We assume that the initial states are initialized either to logical zero states $\ket{0_{M,\Phi}}$, or
states for gate operations via teleportation such as magic states $\ket{T_{M,\Phi}}=\frac{1}{\sqrt{2}}(\ket{0_{M,\Phi}}+e^{i\pi/4}\ket{1_{M,\Phi}})$, plus states $\ket{+_{M,\Phi}}=\frac{1}{\sqrt{2}}(\ket{0_{M,\Phi}}+\ket{1_{M,\Phi}})$, and plus $y$ states $\ket{+i_{M,\Phi}}=\frac{1}{\sqrt{2}}(\ket{0_{M,\Phi}}+i\ket{1_{M,\Phi}})$. We can employ SE to improve state preparation fidelity for these states. We remark that our protocol is also fully compatible with the quantum computation model that does not rely on teleportation for gate operations, e.g., universal quantum computation for cat codes protected by quantum Zeno dynamics~\cite{mirrahimi2014dynamically}. In this case, we need to apply SE only to the logical zero states.

\begin{figure}[t!]
    \centering
    \includegraphics[width=6cm]{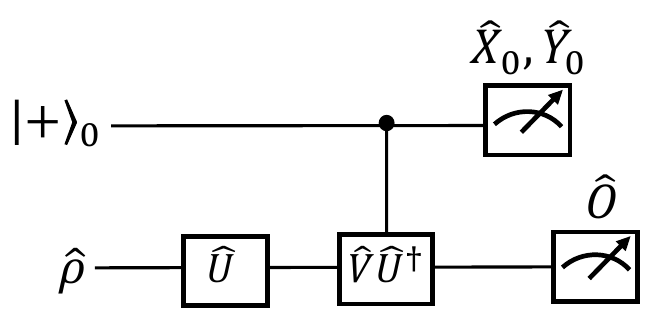}
    \caption{\textcolor{black}{Quantum circuit for measuring Eq. (\ref{Eq: general1}). The black/white circle indicates a control operation that operates when the ancilla qubit is $1/0$. }}
    \label{qcirc2}
\end{figure}

Before we proceed to formulate SE for RSBCs, we review the generalized quantum process introduced in Ref. \cite{sun2022perturbative}.  \textcolor{black}{We consider a quantum circuit for the unitary operators $\hat{U}$ and $\hat{V}$ as shown in Fig. \ref{qcirc2}.} Then, simple calculations show
\begin{equation}
\begin{aligned}
\braket{\hat{X}_0 \otimes \hat{O}}+i\braket{\hat{Y}_0 \otimes \hat{O}}= \tr[\hat{O}  \hat{U} \hat{\rho} \hat{V}^\dag ],
\label{Eq: general1}
\end{aligned}
\end{equation}
where $\hat{X}_0$ and $\hat{Y}_0$ are Pauli operators for the ancilla qubit and $\hat{O}$ is the measured observable. This is equivalent to obtaining the generalized quantum process~\cite{sun2022perturbative}
\begin{equation}
\Psi(\hat{\rho})= \hat{U} \hat{\rho} \hat{V}^\dag.
\label{Eq: generalized}
\end{equation}

\textcolor{black}{Now we use the generalized quantum process for SE for state preparation of RSBCs. The projected state via the projector $\hat{\mathcal{P}}_{\mathbb{S}}$ reads:
\begin{equation}
\begin{aligned}
\hat{\rho}_{{\mathbb{S}}}&= \frac{1}{p_\mathbb{S} |\mathbb{S}|^2} \sum_{\hat{\mathcal{S}}, \hat{\mathcal{S}}' \in \mathbb{S} } \mathcal{S} \hat{\rho} \mathcal{S}'^\dag \\
\label{Eq: initSE}
\end{aligned}.
\end{equation}
Therefore, each generalized process $\hat{\mathcal{S}} (\cdot) \hat{\mathcal{S}}'^\dag$ in Eq. \eqref{Eq: initSE} can be simiulated by replacing the unitaries $\hat{U}$ and $\hat{V}$ in Fig. \ref{qcirc2} with $\mathcal{S}$ and $\mathcal{S}'$. Then, by randomly generating $\mathcal{S}$ and $\mathcal{S}'$, we can simulate the projection by the projector $\hat{\mathcal{P}}_{\mathbb{S}}$ as indicated in Fig. \ref{Fig:largefig1} (c). Note that $\hat{V} \hat{U}^\dag $ in Fig. \ref{qcirc2} is replaced by $\hat{\mathcal{S}} \hat{\mathcal{S}}'^\dag$, which is an element of the group $\mathbb{S}$.  Then, with the normalization of the measurement outcome with the projection probability $p_{\mathbb{S}}$, we can finally apply SE in the state preparation.}

Then, we consider that $\mathbb{S}$ is the set of rotation operators for the error mitigation for rotation symmetric codes. Let us denote the noisy logical zero states as $\hat{\rho}_i$ and the resource states for gate rotation as $\hat{\sigma}_j$, where $i$ and $j$ indicate labels of bosonic qubits. The error-mitigated initial states via SE read
\begin{equation}
\begin{aligned}
\hat{\rho}_{ i}^{\mathrm{S}}&=\frac{1}{p_i} \hat{\mathcal{P}}_{2 M}^{(0)} \hat{\rho}_{i} \hat{\mathcal{P}}_{2 M}^{(0)}=\frac{1}{p_i(2M)^2} \sum_{k_i, k_i'=0}^{2M-1} \hat{Z}_M^{k_i} \hat{\rho}_i  \hat{Z}_M^{k_i'},  \\
\hat{\sigma}_{j}^{\mathrm{S}}&=\frac{1}{q_j} \hat{\mathcal{P}}_{2 M}^{(\rm{c})} \hat{\sigma}_{j} \hat{\mathcal{P}}_{2 M}^{(\rm{c})}=\frac{1}{q_j M^2} \sum_{l_j l_j'=0}^{M-1} \hat{R}_M^{l_j} \hat{\sigma}_j  \hat{R}_M^{l_j'},
\end{aligned}
\end{equation}
where $p_i=\mathrm{Tr}[\hat{\mathcal{P}}_{2 M}^{(0)} \hat{\rho}_{i} ]$ and $q_j=\mathrm{Tr}[\hat{\mathcal{P}}_{2 M}^{(\rm{c})} \hat{\sigma}_{j} ]$ are the projection probabilities.

Let the measured observable and the process in a quantum circuit including the teleportation and error correction procedure denote $O$ and $\mathcal{U}_{\rm C}$ respectively.  We also include the rounding process for classical error correction of the measurement of the observable in $\mathcal{U}_{\rm C}$. Then the error-mitigated expectation value can be written as
\begin{equation}
\begin{aligned}
&\braket{\hat{O}_{\rm SE}}=\mathrm{Tr}[O \mathcal{U}_{\rm C} \big((\bigotimes^{N_{\hat{\rho}}-1}_{i=0} \hat{\rho}^{\rm S}_i) \otimes (\bigotimes^{N_{\hat{\sigma}}-1}_{j=0}\hat{\sigma}_{j}^{\rm S}) \big)] \\
&= \frac{1}{(\prod_i p_i \prod_j q_j) (2M)^{2N_{\hat{\rho}}} M^{2N_{\hat{\sigma}}}} \\
&\times \sum_{\vec{l}, \vec{l}^\prime,\vec{k},\vec{k}^\prime}  \mathrm{Tr}\bigg[\hat{O}  \mathcal{U}_{\mathrm C} \big((\bigotimes_{i=0}^{N_{\hat{\rho}}-1} \hat{Z}_M^{k_{i}} \hat{\rho}_i \hat{Z}_M^{k_{i}^\prime})\otimes (\bigotimes_{j=0}^{N_{\hat{\sigma}}-1} \hat{R}_M^{l_{j}} \hat{\sigma}_j \hat{R}_M^{l_{j}^\prime})  \big)\bigg],
\label{Eq:numerator2}
\end{aligned}
\end{equation}
where $\vec{k}=(k_0,k_1,...,k_{N_{\hat{\rho}}-1})$, $\vec{k}'=(k_0',k_1',...,k_{N_{\hat{\rho}}-1}')$, $\vec{l}=(l_0,l_1,...,l_{N_{\hat{\sigma}}-1})$ and  $\vec{l}'=(l_0',l_1',...,l_{N_{\hat{\sigma}}-1}')$. 
Here, $N_{\hat{\rho}}$
($N_{\hat{\sigma}}$)
denotes the number of bosonic qubits for $\hat{\rho}_i$ ($\hat{\sigma}_i$).
To compute Eq. (\ref{Eq:numerator2}), we first compute the projection probability $p_i$ and $q_j$. The probabilities can be calculated with a classical simulation given the noise model as we will show in Sec. \ref{Sec: performance} or can be straightforwardly evaluated with the linear combination of expectation values of $\hat{Z}_M^k$ or $\hat{R}_M^k$ measured by using a Hadamard test circuit.

To calculate the other part of Eq. (\ref{Eq:numerator2}), i.e., the unbiased estimator for $(\prod_i p_i \prod_j q_j) \braket{\hat{O}_{\rm SE}}$, we first randomly generate $(\vec{k},\vec{k}',\vec{l},\vec{l}')$ with a uniform distribution. Then we use the generalized quantum process to perform unphysical operations $\{\hat{Z}_M^{k_{i}} (\cdot) \hat{Z}_M^{k_{i}'}\}_i$ and $\{\hat{R}_M^{l_{j}} (\cdot) \hat{R}_M^{l_{j}'}\}_j$. To do so, we use one ancilla qubit for each bosonic qubit as shown in Fig. \ref{qcirc3} (a). It is noteworthy that we can recycle the ancilla qubit by turns in each SE procedure. We next perform the process corresponding to the computation $\mathcal{U}_C$, and finally measure the observable $\hat{O}$. We repeat this procedure, and construct the unbiased estimator of $(\prod_i p_i \prod_j q_j) \braket{\hat{O}_{\rm SE}}$.  Notice that because $\braket{\hat{O}_{\rm SE}}, p_i, q_j \in \mathbb{R}$, there is no contribution from the imaginary part of Eq. (\ref{Eq: general1}); hence, it is not necessary to measure Pauli $Y$ operators of ancilla qubits. More concretely, denoting the unbaised estimator of $\braket{\hat{O}_{\rm SE}}$ obtained in this procedure as $\mu_{\rm SE}$, we get
\begin{equation}
\braket{\mu_{\rm SE}}=\frac{1}{(\prod_i p_i \prod_j q_j)}\braket{\vec{X}_0\otimes \hat{O} },
\label{Eq: estimator}
\end{equation}
 where $\vec{X}_0$ is the product of Pauli $X$ operators of ancilla qubits. With respect to the sampling cost of this method, due to the division by $(\prod_i p_i \prod_j q_j)$ in Eq. (\ref{Eq: estimator}), we have
\begin{equation}
\mathrm{Var}[\mu_{\rm SE}] \propto  (\prod_i p_i \prod_j q_j)^{-2} \equiv C,
\label{Eq: amplify}
\end{equation}
which indicates the number of measurements to achieve a certain accuracy $\varepsilon$ scales with $O(( \prod_i p_i \prod_j q_j \varepsilon)^{-2})$.

\textcolor{black}{In our later work \cite{tsubouchi2023virtual}, we generalized the symmetry expansion for initial state preparation so that it can be applied at any point during circuit execution, which is called virtual quantum error detection (VQED). In Ref. \cite{tsubouchi2023virtual}, we considered the Pauli-based stabilizer error correction and detection codes. By replacing the Pauli stabilizers with rotation stabilizer operators, we can mitigate the effect of photon loss for rotation-symmetric codes during circuit execution.}

\subsection{Comparison with error detection for initial state preparation}
We compare our method with \textcolor{black}{the conventional verification method,} i.e., quantum error detection for state preparation. Suppose the case of $M=2^m$ for $m \in \mathbb{N}$. For performing verification of the logical zero state $\ket{0_{M,\Phi}}$ via the conventional quantum error detection using an ancilla qubit, we need to perform Hadamard-test gates for rotation unitaries $\{e^{i 2\pi/2^k} \}_{k=1}^{m+1}$ and post-select the verified states depending on the measurement outcomes of the ancilla qubit~\cite{grimsmo2020quantum}. \textcolor{black}{Therefore, it is necessary to perform $m+1$ Hadamard-test gates with high-fidelity
single-shot measurements of the ancilla qubit.}

Meanwhile, our method only requires a constant number of control operations by dispersive interactions and measurement of expectation values of Pauli $X$ operators of ancilla qubits. 
Therefore, when high-fidelity single-shot measurements are not available,
our proposal will clearly outperform the conventional verification protocol for large $M$. \textcolor{black}{Furthermore, our method is fully compatible with readout error mitigation methods~\cite{maciejewski2020mitigation,bravyi2021mitigating} for the readout of ancilla qubit, which can dramatically improve the computation accuracy, considering the fact that the readout fidelity is relatively lower than gate operation fidelity for qubit-based computation.}

For a fair comparison, we mention the downsides of our protocol. Regarding the sampling cost, while the verification succeeds with the probability $\prod_ip_i \prod_j q_j$, the number of measurements required in our protocol scales with $O((\prod_ip_i \prod_j q_j)^{-2})$, which indicates quadratically worse sampling overheads. Another disadvantage is that quantum error mitigation, including SE, is restricted to algorithms relying on expectation values of observables. This problem may not be so significant because most NISQ algorithms, e.g., the variational quantum eigensolver (VQE)~\cite{peruzzo2014variational,tilly2022variational,bharti2022noisy,cerezo2021variational}, employs only expectation values. It has also been shown that quantum error mitigation can be incorporated in phase estimation algorithms~\cite{suzuki2022quantum}. 

\subsection{Symmetry expansion before measurement}
Next, we introduce the method in which SE is applied immediately before measurement. Although this scenario is discussed for Pauli-stabilizer symmetries, we propose a method tailored to RSBCs. Suppose that 
\textcolor{black}{we cannot use the information of the symmetry
for error correction through a single-shot measurement}
due to the low resolution of measurement outcomes, e.g., photon parity measurement is difficult. In this case, since the measurement fidelity is accordingly low, we consider using a Hadamard test circuit to measure the observable in the logical $Z$ basis. In this case, we have $\hat{O}=\hat{Z}_M^{\otimes N_{\rm M}}$, where $ N_{\rm M}$ is the number of measured bosonic qubits.  The error-mitigated expectation value of an observable results in 
\begin{equation}
\braket{\hat{O}}_{\rm{SE}}= \frac{\mathrm{Tr}[(\hat{Z}_M \hat{\mathcal{P}}_{ 2M}^{(\mathrm{c})})^{ \otimes N_{\mathrm M}} \hat{\rho}_{\rm noisy} ]}{\mathrm{Tr}[( \hat{\mathcal{P}}_{ 2M}^{(\mathrm{c})})^{ \otimes N_{\mathrm M}} \hat{\rho}_{\rm noisy} ]},
\label{Eq: EqQEM}
\end{equation}
where $N_{\mathrm{M}}$ is the number of measured qubits in the observable and $\hat{\rho}_{\rm noisy}$ is the noisy output state.

The numerator can be expanded as
\begin{equation}
\begin{aligned}
&\mathrm{Tr}[(\hat{Z}_M \hat{\mathcal{P}}_{ 2M}^{(\mathrm{c})})^{ \otimes N_{\mathrm M}} \hat{\rho}_{\rm noisy} ]\\
&= \frac{1}{ M^{2 N_{\rm M}} }  \sum_{m_0,... ,m_{N_{\mathrm{M}}-1}=0}^{M-1} \mathrm{Tr}[ \bigotimes^{N_{\rm M}-1}_{h=0} \hat{Z}_M^{2 m_h+1} \hat{\rho}_{\rm noisy}].
\label{Eqnumerator}
\end{aligned}
\end{equation}

\begin{figure*}[t!]
    \centering
    \includegraphics[width=1.7\columnwidth]{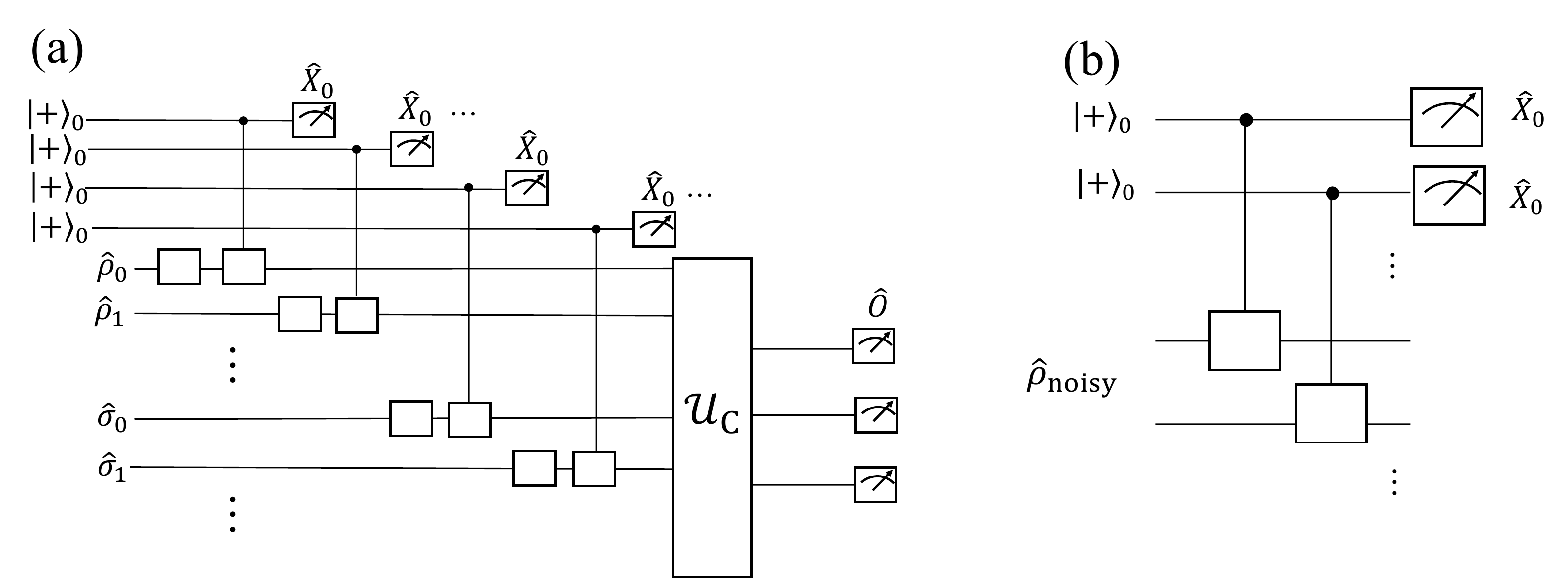}
    \caption{Quantum circuits for performing SE for (a) state preparation and (b) before measurement. Empty squares in the circuits denote rotation operations.}
    \label{qcirc3}
\end{figure*}

Eq. (\ref{Eqnumerator}) can be evaluated by randomly generating $(m_0, m_1, ..., m_{N_{\mathrm{M}}-1})$ with a uniform distribution, and measuring ancilla qubits in the quantum circuit shown in Fig. \ref{qcirc3} (b). We only need Pauli $X$ measurements of ancilla qubits because $\mathrm{Tr}[(\hat{Z}_M \hat{\mathcal{P}}_{2 M}^{(\mathrm{c})})^{ \otimes N_{\mathrm M}} \hat{\rho}_{\rm noisy} ]$ is real. The denominator can also be calculated similarly, which corresponds to the projection probability of the noisy state onto the symmetric subspace. Denoting the denominator as $p_{\rm c}$, the sampling cost of this method is proportional to $p_{\rm c}^{-2}$ with the same argument in Sec \ref{Sec: SE}.

We remark that we can straightforwardly combine SE for state preparation and measurement for further enhancement of computation results. The sampling overhead grows with $O((p_{\rm c} (\prod_i p_i \prod_j q_j)  \varepsilon)^{-2})$ for a required accuracy $\varepsilon$.

\subsection{Virtual creation of RSBC states}
As a remarkable application of SE, we can virtually perform state preparation of RSBCs only from a primitive state $\hat{\rho}_\Phi=\ket{\Phi}\bra{\Phi}$ with the same setup for error mitigation. For $\hat{\rho}_{M,\Phi}^{(0)}=\ket{0_{M,\Phi}}\bra{0_{M,\Phi}}$, denoting $p_\Phi=\mathrm{Tr}[\hat{\mathcal{P}}_{2M}^{(0)} \hat{\rho}_\Phi]$, since
\begin{equation}
\hat{\rho}_{M,\Phi}^{(0)}=\frac{1}{p_\Phi} \hat{\mathcal{P}}_{2M}^{(0)} \hat{\rho}_\Phi \hat{\mathcal{P}}_{2M}^{(0)}
\end{equation}
we can obtain the expectation value corresponding to the state $\hat{\rho}_{M,\Phi}^{(0)}$ with the same implementation of SE. For primitive states such that $\bra{\Phi} \hat{Z}_M^k \ket{\Phi}=\delta_{k,0}$, e.g., coherent states with a sufficiently large photon count, we have $p_\Phi=1/2M$. This indicates that the sampling overhead increases as $O((2M)^{2 N_{\rm v}})$, where $N_{\rm v}$ is the number of virtual bosonic qubits; therefore, although we cannot rely on this method for state preparation of all the bosonic qubits, when experimentalists are short of a couple of bosonic qubits, this method may be able to compensate for that shortage.

\textcolor{black}{\section{Phase-error suppression}}
Here, we discuss the mitigation of phase error. The number phase codes have other symmetries described by the logical $X$ operator $\hat{X}_M=\sum_{m=0}^\infty \ket{m}\bra{m+M}$. Since $\hat{X}_M^{2k} ~(k\in\mathbb{N})$ acts as a logical identity operator and therefore is a stabilizer on the code space relevant to phase error suppression, we can define the corresponding projector $\hat{\mathcal{P}}_X \propto \sum_{k=0}^\infty  \hat{X}_M^{2k}$. See Appendix. \ref{sec:supplec} for the performance of the truncated projectors $\hat{\mathcal{P}}_X^{(L)} \propto \frac{1}{L+1}\sum_{k=0}^{L}  \hat{X}_M^{2k}~(L\in \mathbb{N})$  for mitigating phase errors. However, it is not realistic to construct the projector $\hat{\mathcal{P}}_X^{(L)}$ due to the difficulty of implementing the number-translation operator.

Alternatively, we propose a method to suppress phase errors by developing the recently proposed projective squeezing method~\cite{endo2024projective,anai2024unitary}. The intuition behind the choice of the projective squeezing method is that this method allows for the projection onto a squeezed vacuum state with a higher squeezing level, and the squeezed vacuum state with a higher squeezing level has a smaller phase uncertainty. Then, by considering the rotation symmetries as well, we can construct the projector onto the rotation codes with small phase uncertainty. Here, we discuss the suppression of the phase error by using the example of four-legged rotation-symmetric codes. We consider the four-legged rotation code with the primitive state being the squeezed vacuum states for the squeezing parameter $r>0$:
\begin{equation}
\begin{aligned}
&\ket{\pm_{M=2}, r}= \hat{S}(\mp r) \ket{0}.  \\
\label{Eq: phasesqstate}
\end{aligned}
\end{equation}
Note that the code states described by Eq. \eqref{Eq: phasesqstate} have increasingly small phase uncertainty as $r$ increases. To construct the projector onto the code space described by Eq. \eqref{Eq: phasesqstate}, we develop the projective squeezing method introduced in Refs. \cite{endo2024projective, anai2024unitary}. The smeared projector to increase the squeezing level from $r$ to $r+ \Delta r$ for $\ket{-_{M=2}, r}$ is described by
\begin{equation}
\begin{aligned}
\tilde{P}_{\rm sq} (\Gamma)&= \int ds \sqrt{\frac{\Gamma}{\pi}} e^{-\Gamma s^2} \hat{D}\left( i \frac{s}{\sqrt{2}}\right), \\
&= \mathrm{exp}\left(-\frac{\hat{x}^2}{4 \Gamma} \right)
\end{aligned}
\end{equation}
which increases the squeezing level by $\Delta r$ with $\Delta r = \frac{1}{2} \mathrm{ln} \left(1+\frac {1}{2} \Gamma^{-1} e^{-2 r} \right)$. Here, $\hat{D}(\alpha)~\alpha \in \mathbb{C}$ is a displacement operator. Meanwhile, by changing $\hat{D}\left(i \frac{s}{\sqrt{2}} \right)$ with $\hat{D}\left( \frac{s}{\sqrt{2}} \right)$, we can construct the smeared projector for the four-legged rotation code that squeezes the state for the momentum axis $\tilde{P}_{\rm asq} (\Gamma)= \mathrm{exp}\left(-\frac{\hat{p}^2}{4 \Gamma} \right)$. Now, we construct the smeared projector to mitigate the phase error as follows: 
\begin{equation}
\tilde{P}_{\rm phase}(\Gamma)= \tilde{P}_{\rm sq} (\Gamma) +  \tilde{P}_{\rm asq} (\Gamma).
\end{equation}
We can show 
\begin{equation}
\begin{aligned}
&\tilde{P}_{\rm phase}(\Gamma) (c_+ \ket{+_{M=2}, r}+ c_- \ket{-_{M=2}, r}) \\
&\sim e^{- \frac{\Delta r}{2} } (c_+ \ket{+_{M=2}, r+\Delta r} + c_- \ket{-_{M=2}, r+\Delta r}),
\end{aligned}
\end{equation}
which indicates that the phase uncertainty is suppressed. Here, the norms of $\tilde{P}_{\rm sq}(\Gamma) \ket{+_{M=2},r}$ and $\tilde{P}_{\rm asq}(\Gamma) \ket{-_{M=2},r}$ are exponentially small with the squeezing level $r$, i.e.,
\begin{equation}
\begin{aligned}
&\|\tilde{P}_{\rm sq}(\Gamma) \ket{+_{M=2},r} \|^2 = \|\tilde{P}_{\rm asq}(\Gamma) \ket{-_{M=2},r} \|^2 \\
&= \frac{e^{-2r}}{\sqrt{ e^{2 \Delta r} -1+e^{-3r}} },
\end{aligned}
\end{equation}
and we ignore these terms.

We numerically demonstrate that the smeared projector $\tilde{P}_{\rm phase}(\Gamma)$ reduces phase uncertainty, thereby mitigating phase errors for four-legged cat states, both in the pure state case and under phase errors described by a Lindblad operator $\hat{a}^\dag \hat{a}$.

Now, we discuss the implementation and the sampling cost for applying the smeared projector $\tilde{P}_{\rm phase}(\Gamma)$. Because we can expand the normalized smeared projector $\tilde{P}'_{\rm phase} (\Gamma) =\frac{1}{2} \tilde{P}_{\rm phase} (\Gamma) $ as: 
\begin{equation}
\begin{aligned}
&\tilde{P}'_{\rm phase} \\
&= \int ds \frac{1}{2}\sqrt{\frac{\Gamma}{\pi}} e^{-\Gamma s^2} \hat{D}\left( i \frac{s}{\sqrt{2}}\right) +  \int ds \frac{1}{2}\sqrt{\frac{\Gamma}{\pi}} e^{-\Gamma s^2} \hat{D}\left( \frac{s}{\sqrt{2}}\right) \\
&\sim \sum_k p_k \hat{D}_k,
\end{aligned}
\end{equation}
where $\{ \hat{D}_k \}_k$ corresponds to either $\{ \hat{D}\left( i \frac{s}{\sqrt{2}}\right)\}_s$ or $\{ \hat{D}\left(\frac{s}{\sqrt{2}}\right)\}_s$ with the approximation being necessary for the proper discretization for the implementation. Note that $\sum_k p_k=1$ due to the normalization. While a detailed explanation of the implementation of the projective squeezing can be found in Refs. \cite{endo2024projective, anai2024unitary}, this can be considered a generalized method of the symmetry expansion method. We randomly generate the displacement operations with the probability of $\{p_k \}_k$. Then, by replacing the controlled rotation operations with the controlled displacement operations, we can perform the virtual projection due to the smeared projector $\tilde{P}_{\rm phase} (\Gamma)$. Now, the sampling cost is described by the projection probability $q_{\rm phase} (\Gamma) = \mathrm{Tr}[\tilde{P}_{\rm phase}'^\dag \tilde{P}_{\rm phase}' \hat{\rho}_{\rm in}]$ for the input state $\hat{\rho}_{\rm in}$. When the input state is $c_+ \ket{+_{M=2}, r}+ c_- \ket{-_{M=2}, r}$, the projection probability reads:
\begin{equation}
q_{\rm phase} (\Gamma) \sim \frac{1}{4} e^{-\Delta r}, 
\label{Eq: projectionphase}
\end{equation}
where the coefficient $1/4$ appears as a result of the normalization, which leads to a large sampling cost, i.e., the sampling overhead in Eq. \eqref{Eq: amplify} scales as 
\begin{equation}
C_{\rm phase} \sim 16 e^{2 \Delta r}. 
\end{equation}

\textcolor{black}{\section{Robust implementation of SE against ancilla-qubit errors}}
Here, we review the result in Ref. \cite{endo2024projective}, which offers a robust implementation of our framework to the ancilla qubit errors. We model the noisy interaction for controlled operations between the ancilla qubit and the resonator via the following Lindblad master equation:
\begin{equation}
\frac{d}{dt}\hat{\rho}(t)= -i[\hat{Z} \otimes \hat{V}', \hat{\rho}(t)] + w_{1} \mathcal{D}[\hat{\sigma}^{-1}] (\hat{\rho}(t))+w_{2} \mathcal{D}[\hat{Z}] (\hat{\rho}(t)),
\label{Eq: Lind}
\end{equation}
where $\mathcal{D}[\hat{A}] (\hat{\rho})= \frac{1}{2}(2 \hat{A} \hat{\rho} \hat{A}^\dag - \hat{A}^\dag \hat{A} \hat{\rho}- \hat{\rho}  \hat{A}^\dag \hat{A} )$ is the Lindblad superoperator, $w_1$ and $w_2$ are $T_1$ and $T_2$ error rates, and $Z \otimes V'$ is the interaction Hamiltonian for realizing the controlled operation. Introducing the block representation of the density matrix of the composite system
\begin{equation}
\hat{\rho}_{\rm com}=\begin{pmatrix}
\hat{\rho}_{00}(t) & \hat{\rho}_{01} (t) \\
\hat{\rho}_{10} (t) & \hat{\rho}_{11} (t)
\end{pmatrix},
\end{equation}
where the subindices correspond to the ancilla qubit state. Then, by comparing both sides of Eq. \eqref{Eq: Lind}, we can show:
\begin{equation}
\begin{aligned}
\frac{d \hat{\rho}_{01}}{dt} &= -i \{V', \hat{\rho}_{01} \}- \bigg( \frac{w_1}{2} + 2 w_2 \bigg) \hat{\rho}_{01} \\
\frac{d \hat{\rho}_{10}}{dt} &= i \{V', \hat{\rho}_{10} \}- \bigg( \frac{w_1}{2} + 2 w_2 \bigg) \hat{\rho}_{10}.
\end{aligned}
\end{equation}

and hence

\begin{equation}
\begin{aligned}
\hat{\rho}_{01}(t)&= e^{-(\frac{w_1}{2}+2w_2) t} \hat{\rho}_{01}^{\rm id}(t) \\
\hat{\rho}_{10}(t)&= e^{-(\frac{w_1}{2}+2w_2) t} \hat{\rho}_{10}^{\rm id}(t).
\end{aligned}
\end{equation}
Here, $\hat{\rho}_{10 (01)}^{\rm id}(t)$ is the noiseless block matrix whereas $\hat{\rho}_{10 (01)}(t)$ is the noisy one. Measuring the Pauli X operator of the ancilla qubit yields:
\begin{equation}
\begin{aligned}
\mathrm{Tr}[\hat{\rho}_{\rm com} (t) X \otimes I] &= \hat{\rho}_{01} (t) + \hat{\rho}_{10} (t) \\
&= e^{-(\frac{w_1}{2} + 2w_2) t} (\hat{\rho}_{01}^{\rm id} (t)+ \hat{\rho}_{10}^{\rm id} (t) ), \\
\mathrm{Tr}[\hat{\rho}_{\rm com} (t) X \otimes O] &=\mathrm{Tr}[ (\hat{\rho}_{01} (t) + \hat{\rho}_{10} (t)) O] \\
&= e^{-(\frac{w_1}{2} + 2w_2) t} \mathrm{Tr}[(\hat{\rho}_{01}^{\rm id} (t)+ \hat{\rho}_{10}^{\rm id} (t) ) O]. 
\label{Eq: uniformdecay}
\end{aligned}
\end{equation}
 Eq. \eqref{Eq: uniformdecay} shows that $T_1$ and $T_2$ errors only uniformly shrink the expectation value when we measure the Pauli X of the ancilla qubit.

Now, because our protocol uses the same circuit for computing the numerator and denominator in Eq. \eqref{Eq:numerator2} except for the measurement of the observable, the effects of the noise are likely to be canceled out. Furthermore, we can construct the unbiased estimator of the error-mitigated expectation value by changing the sampling probability of error-mitigation operations~\cite{endo2024projective}. We denote the generalized processes under noise as 
\begin{equation}
\Psi'_k(\rho)= r_k \hat{U}_k \hat{\rho} \hat{V}_k^\dag,
\end{equation}
where $0< r_k <1 $ is a coefficient due to noise. While the uniform linear combination of noiseless generalized processes $\Psi_k$ realizes a projection onto the rotation code subspace due to the projector $\hat{\mathcal{P}}$, we obtain
\begin{equation}
\frac{1}{c \mathcal{M}} \sum_k \Psi_k =  \sum_k q_k \Psi'_k,
\end{equation}
where $\mathcal{M}$ is the number of generalized processes for QEM, $c= \frac{1}{\mathcal{M}} \sum_k r_k^{-1} >1$ and $q_k = \frac{1}{\mathcal{M} c r_k}$ with $\sum_k q_k =1$. For non-uniform linear combinations of generalized processes for projective squeezing, a similar modification is possible~\cite{endo2024projective}. If we estimate the projection probability with random sampling of Hadamard test circuits in a similar vein to the estimation of the numerator in Eq. \eqref{Eq: estimator}, the effect of the rescaling factor $c$ is canceled out due to division by modifying the sampling probability of the generalized process for QEM to $\{ q_k\}_k$. Thus, we can construct the unbiased estimator of the error-mitigated expectation values even under ancilla-qubit errors. Note that the sampling overhead is amplified with $c^2$ because the denominator is rescaled by $c^{-1}$. 

\indent

\section{Performance evaluation}
\label{Sec: performance}
In this section, we evaluate the performance of SE for cat codes and binomial codes from analytical and numerical studies.
We here consider photon loss noise $\frac{d \hat{\rho} }{dt} = \frac{\gamma}{2} (2a \hat{\rho} a^\dag -a^\dag a \hat{\rho} -\hat{\rho} a^\dag a)$.  Refer to Appendix~\ref{Appendix: A} for the detailed analytical derivation of the results.
 \subsection{Symmetry expansion for logical zero cat states}
\begin{figure}[t!]
    \centering
    \includegraphics[width=\columnwidth]{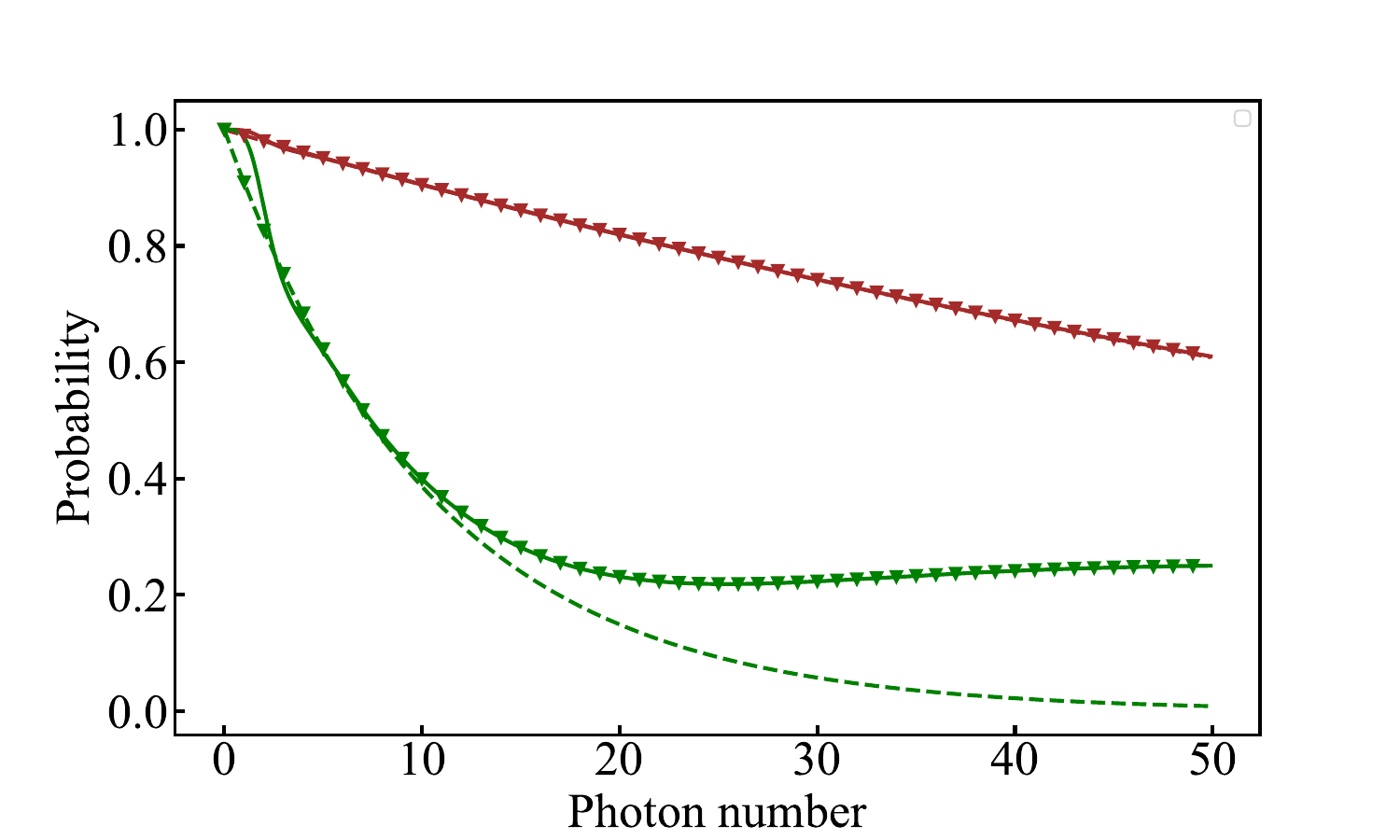}
    \caption{Projection probability $p_0$ versus the photon number $|\alpha|^2$ for $M=2$. Brown and green indicate the low-error and high-error regimes $\gamma t=0.01$ and $\gamma t=0.1$, respectively. Lined curves, triangles, and dashed curves correspond to the exact analytical solution Eq. (\ref{Eq: projectionpro1}), approximations in Eq. (\ref{Eq: p02}) and Eq. (\ref{Eq: p03}).  }
    \label{Figzero}
\end{figure}

 Here, we discuss the performance of SE with analytical studies
 for cat-type RSBCs where the primitive state is a coherent state, i.e., $\ket{\Phi}=\ket{\alpha}$. When we consider a logical $0$ state $\ket{0_{M,\alpha}}$, the state at time $t$ reads:
 \begin{equation}
 \begin{aligned}
\hat{\rho}(t) = \frac{e^{-\Gamma (t)}}{\mathcal{N}}\sum_{m, m'=0}^{2M-1}  e^{\Gamma (t) e^{i\frac{m-m'}{M}}\pi} \ket{\alpha (t) e^{i\frac{m}{M} \pi}  } \bra{\alpha (t) e^{i\frac{m'}{M} \pi}  },
\end{aligned}
 \end{equation}
 where $\mathcal{N}$ is the normalization factor, $\Gamma (t) = \alpha^2 (1-e^{-\gamma t})$ is the effective error rate, and $\alpha(t)=\alpha e^{-\gamma t/2}$. Note that we approximately have $\Gamma(t) \sim \alpha^2 \gamma t$ for $\gamma t \ll 1$. Now, with symmetry expansion, we can obtain
 \begin{equation}
\hat{\mathcal{P}}_{2 M}^{(0)} \hat{\rho}(t) \hat{\mathcal{P}}_{2 M}^{(0)} \propto \ket{0_{M, \alpha (t)}}\bra{0_{M, \alpha (t)}}.
 \end{equation}
Here, we use $\hat{\mathcal{P}}_{2M}^0 \ket{\alpha (t) e^{i\frac{m}{M} \pi} } \propto \ket{0_{M, \alpha (t) }}~ \forall m$. Thus, we can see that the state is perfectly projected onto the symmetric subspace corresponding to the primitive state $\ket{\Phi}=\ket{ \alpha (t)}$. Accordingly, defining $f_M(x)= \sum_{l=0} \frac{(x)^{Ml}}{(Ml)!}$, the exact analytical form of the projection probability can be obtained as
\begin{equation}
\begin{aligned}
p_0(t)=\mathrm{Tr}[\hat{\mathcal{P}}_{2M}^0 \hat{\rho}(t) ]=\frac{f_{2M}(\Gamma (t)) f_{2M}(\alpha (t)^2)}{f_{2M}(\alpha^2)}.
\label{Eq: projectionpro1}
\end{aligned}
\end{equation}
Eq. (\ref{Eq: projectionpro1}) can be approximated as
\begin{equation}
p_0(t) \sim e^{-\Gamma (t)} f_{2M} (\Gamma (t))
\label{Eq: p02}
\end{equation}
in the regime where the photon count is large enough and the error rate is low as well.
We note that, if $\Gamma (t) \lesssim 1$, $p_0(t)$ can be further approximated as 
\begin{equation}
p_0(t) \sim e^{-\Gamma (t)}.
\label{Eq: p03}
\end{equation}
In the limit of $\alpha \rightarrow \infty,$ we get $p_0(t) \rightarrow 1/2M $ in \eqref{Eq: projectionpro1}. In Fig. \ref{Figzero}, we plot the projection probability $p_0$ in the low-error and high-error regimes for $\gamma t= 0.01$ and $\gamma t=0.1$. We can observe that in the low-error regime, the approximations agree well with the exact analytical solution Eq. \eqref{Eq: projectionpro1}. In the large-error regime, we can see the behavior until the convergence of the projection probability to $1/2M$. As the photon number increases, the approximation of Eq. (\ref{Eq: p03}) breaks down because $\Gamma (t) \lesssim 1$ is not satisfied. Instead, Eq. (\ref{Eq: p02}) well-captures the dynamics because terms in the series of $f_{2M} (\Gamma (t))$ are non-negligible. We can also observe that in the low photon number regime, the projection probability stays closer to unity in the exact solution; this is because states adjacent to the vacuum state are highly symmetric.

 \subsection{Symmetry expansion for general logical cat states}
\begin{figure}[h!]
    \centering
    \includegraphics[width=\columnwidth]{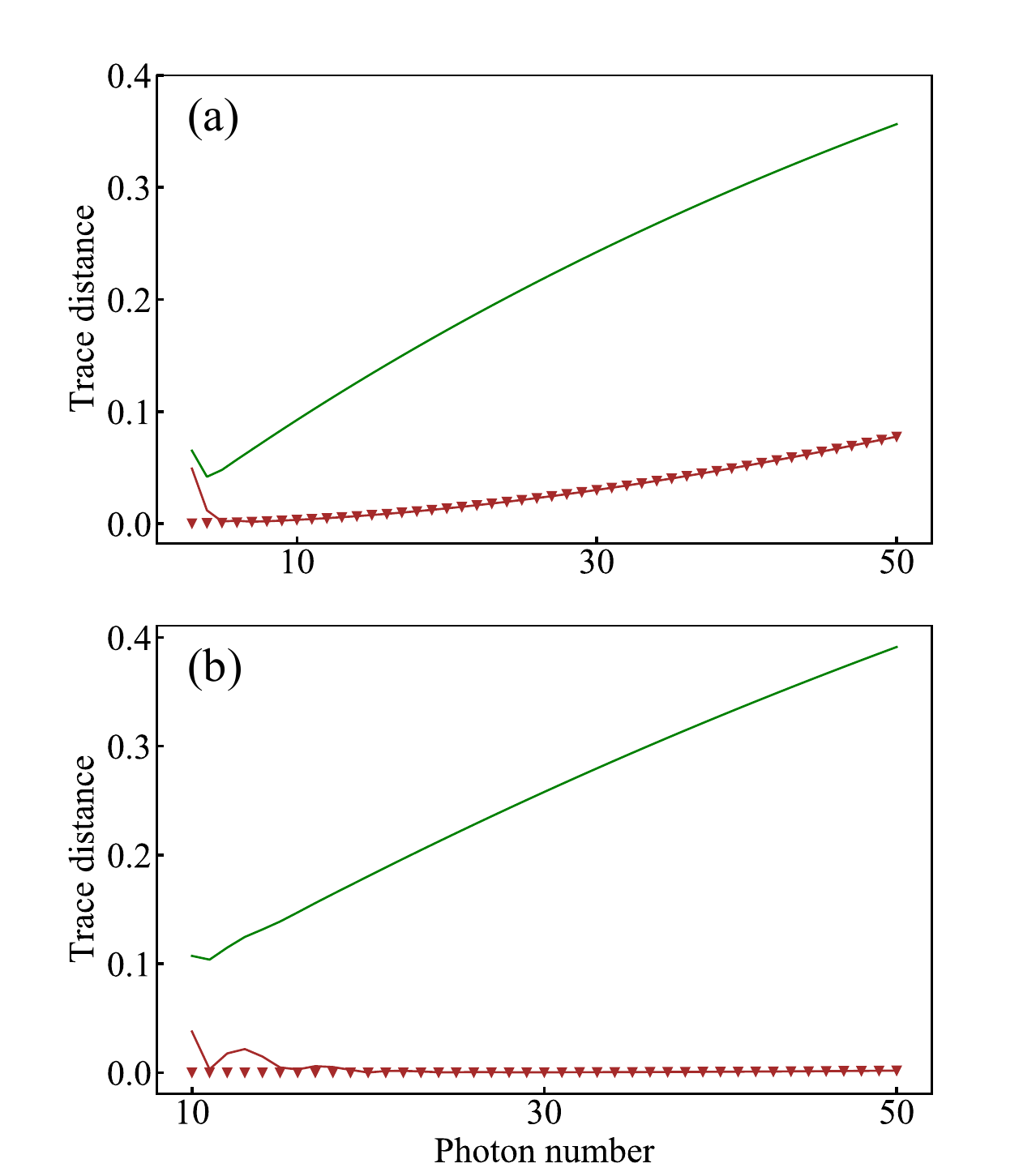}
    \caption{Trace distance between the state with/without error mitigation and the rotation code state $\hat{\rho}_{T,\alpha(t)}$ depending on the photon number $|\alpha|^2$. Green/Brown lined curves indicate the exact numerical results for the unmitigated/error-mitigated quantum states. Brown triangles are the analytical results given in Eq. (\ref{Eq: tracedistance}). We set (a) $M=2$ and (b) $M=4$. }
    \label{Figtracedist}
\end{figure}

For symmetry expansion of general logical states $\ket{\psi_{M,\alpha}}=a \ket{0_{M, \alpha}}+b \ket{1_{M, \alpha}}$, we apply the projector onto the code space $\hat{\mathcal{P}}_{2M}^{(\mathrm{c})}$. Let the initial state denote
 \begin{equation}
\hat{\rho}_\psi (0)=\sum_{m, m'=0}^{2M-1} c_m c_{m'}^* \ket{\alpha e^{i\frac{\pi m}{M}}}\bra{\alpha e^{i\frac{\pi m'}{M}}}.
 \end{equation}
Here, $c_{m}=\frac{a}{\sqrt{\mathcal{N}_0}}+(-1)^m \frac{b}{\sqrt{\mathcal{N}_1}}$ and $\mathcal{N}_0$ and  $\mathcal{N}_1$ are the normalization factors of $\ket{0_{M, \alpha}}$ and $\ket{1_{M, \alpha}}$. Then, the state after time $t$ reads:

 \begin{equation}
 \begin{aligned}
\hat{\rho}_\psi(t)&=\sum_{m, m'=0}^{2M-1} c_m c_{m'}^* e^{\Gamma (t)(e^{i\frac{m-m'}{M} \pi} -1)}  \\
&\times  \ket{\alpha (t) e^{i\frac{m}{M} \pi} } \bra{\alpha (t) e^{i\frac{m'}{M} \pi}  }.
\end{aligned}
 \end{equation}
 Then application of the projector $\hat{\mathcal{P}}_{2M}^{(\mathrm{c})}$ and the normalization gives
 \begin{equation}
\begin{aligned}
&\hat{\rho}_\psi^{\rm EM}(t) \\
&=\frac{1}{\mathcal{N}_c}\bigg( |c_0|^2\ket{+(t)}\bra{+(t)}+{c_0 c_1^*}\frac{g_M(\Gamma (t)) }{f_M(\Gamma (t)) }\ket{+(t)}\bra{-(t)} \\
&+{c_0^* c_1}\frac{g_M(\Gamma (t)) }{f_M(\Gamma (t))}\ket{-(t)}\bra{+(t)} +{|c_1|^2}\ket{-(t)}\bra{-(t)}\bigg),
\end{aligned}
 \end{equation}
 
 where $\mathcal{N}_c=|c_0|^2+|c_1|^2$, $\ket{+(t)}=  \frac{1}{\sqrt{M}} \sum_{k=0}^{M-1} e^{i \frac{2\pi k}{M} \hat{N}} \ket{\alpha(t)} \sim \frac{1}{\sqrt{2}} (\ket{0_{M, \alpha(t)}}+ \ket{1_{M, \alpha(t)}})$,  $\ket{-(t)}=  \frac{1}{\sqrt{M}} \sum_{k=0}^{M-1} e^{i \frac{2\pi k}{M} \hat{N}} \ket{e^{i\frac{\pi}{M}} \alpha (t)} \sim \frac{1}{\sqrt{2}} (\ket{0_{M, \alpha (t)}}- \ket{1_{M, \alpha (t)}})$, and $g_M(x)= \sum_{l=0} \frac{x^{Ml}}{(Ml)!}(-1)^l$. The trace distance between $\hat{\rho}_\psi^{\rm EM}(t)$ and the state $\ket{\psi_{M, \alpha(t)}}$, which is denoted as $\hat{\rho}_{\psi, \alpha(t)}$, can be written as
 \begin{equation}
D(\hat{\rho}_\psi^{\rm EM}(t), \hat{\rho}_{\psi, \alpha (t)})=\frac{\mathcal{M}_c}{\mathcal{N}_c} \bigg(1- \frac{g_M(\Gamma(t)) )}{f_M(\Gamma (t) )}\bigg), 
\label{Eq: tracedist}
 \end{equation}
with $\mathcal{M}_c=|c_0| |c_1|$. Taylor-expanding $\frac{1}{f_M(x)}=1-\frac{x^M}{M!}+O(x^{2M})$, 
\textcolor{black}{we obtain} 
\begin{equation}
\begin{aligned}
D(\hat{\rho}_\psi^{\rm EM}(t), \hat{\rho}_{\psi, \alpha (t) }) = \frac{2 \mathcal{M}_c}{\mathcal{N}_c} \frac{\Gamma(t)^M}{M!} +O(\Gamma(t)^{2M}).
\label{Eq: tracedistance}
\end{aligned}
\end{equation}
This clearly shows that the lower order of the error is eliminated as the order of symmetry increases for $\Gamma(t) \lesssim 1$. We plot the trace distance between the ideal state and the noisy/error mitigated states for the magic state $\ket{T_{M,\alpha}}=\frac{1}{\sqrt{2}}(\ket{0_{M,\alpha}}+e^{i \pi/4}\ket{1_{M,\alpha}})$ in Fig. \ref{Figtracedist} for $M=2$ and $M=4$ in the regime of $\gamma t =0.01$. Because our analysis is valid
when coherent states constituting the logical states are sufficiently distinguishable, we restrict ourselves to such a regime in the plot. Refer to Appendix~\ref{Appendix: B} for details. 

\begin{figure}[t!]
    \centering
    \includegraphics[width=\columnwidth]{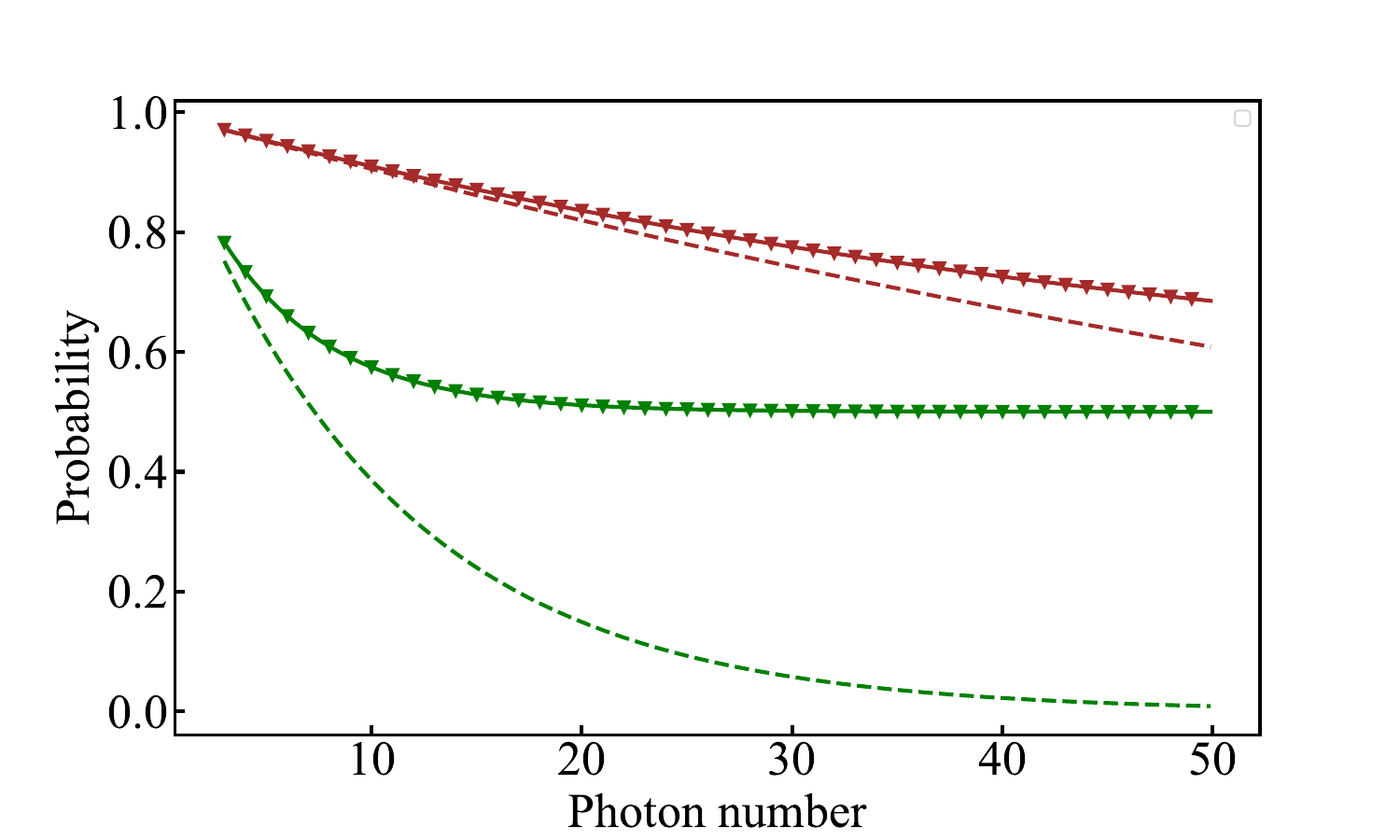}
    \caption{Projection probability $p_T$ versus the photon number $|\alpha|^2$ for $M=2$. Brown and green indicate the low-error and high-error regimes for $\gamma t=0.01$ and $\gamma t=0.1$, respectively. Lined curves, triangles and dashed curves correspond to the exact numerically simulated result, an approximation of Eq. (\ref{Eq: pt}), and  an approximation of $p_T \sim e^{-\Gamma (t)}$.
     }
    \label{prot2}
\end{figure}

Meanwhile, the corresponding projection probability is
 \begin{equation}
\begin{aligned}
 p_\psi& \sim e^{-\Gamma (t)} f_M(\Gamma (t)),
 \label{Eq: pt}
\end{aligned}
 \end{equation}
where we assume $\braket{+(t)|-(t)}\sim 0$. We remark that the projection probability $p_\psi$ can be further approximated to be $p_\psi \sim e^{-\Gamma (t)}$ in the case of $\Gamma (t) \lesssim 1$. We also show the projection probabilities for the noisy magic state onto the code space via SE for $M=2$ in Fig. \ref{prot2} in the regime where each coherent state of the cat code is sufficiently distinguishable. Note that Eq. (\ref{Eq: pt}) agrees well with the exact result for $\gamma t=0.01$ and $\gamma t=0.1$ until the convergence while the approximation by $p_\psi \sim e^{- \Gamma (t)}$ fails to explain the dynamics. This is because the contribution of $f_M(\Gamma(t))$ is non-negligible over the regime. 

\subsection{Numerical simulation for the binomial code}
\begin{figure}[h!]
    \centering
    \includegraphics[width=\columnwidth]{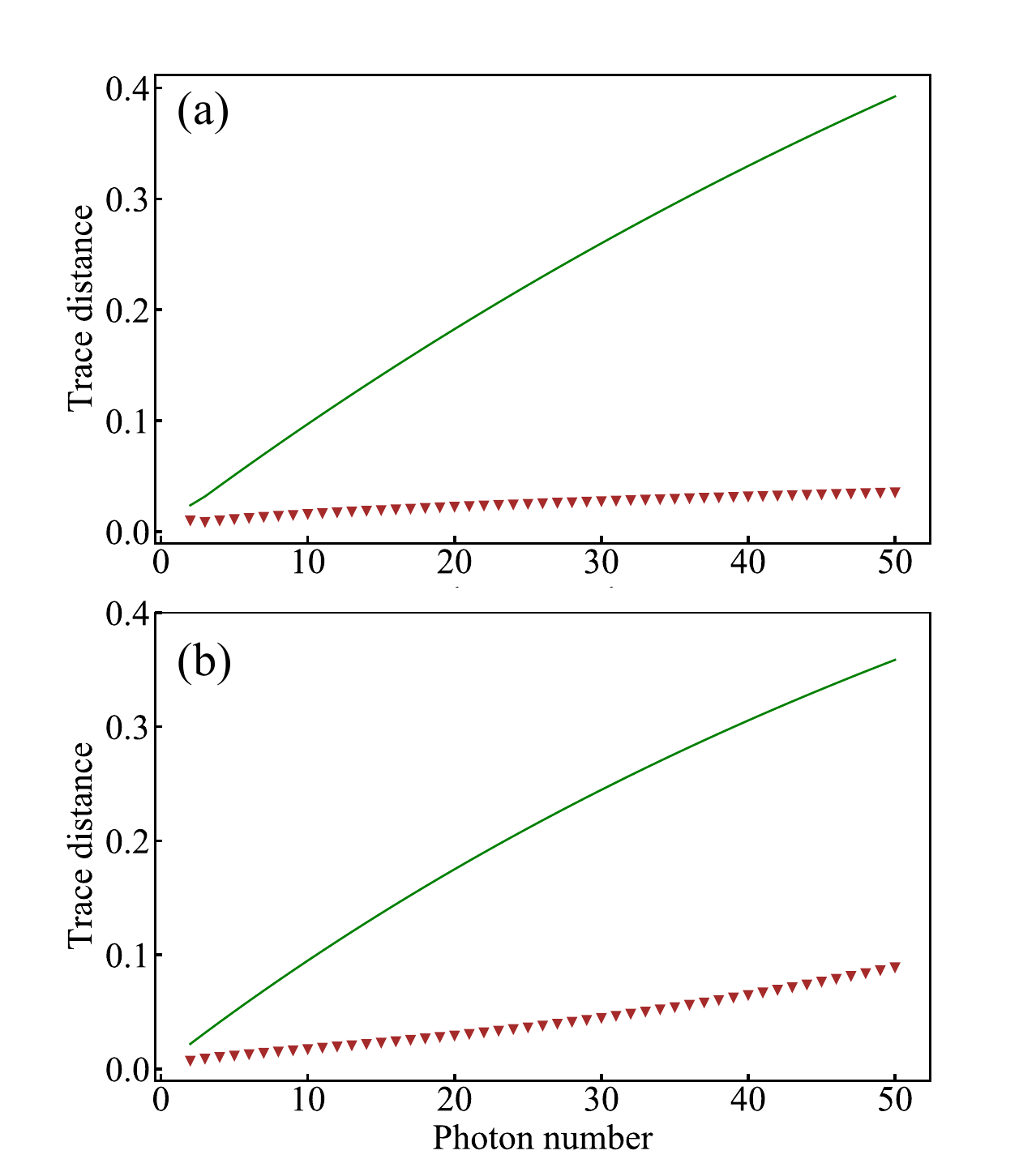}
    \caption{Trace distance between the unmitigated/ error-mitigated states and the error-free states against the photon number for the binomial code for (a) the logical zero states and (b) the logical magic states. We set $M=2$.  
    Green-lined curves correspond to the unmitigated case with the brown triangles denoting the error-mitigated case.}
    \label{Figtracedistbinom}
\end{figure}

\begin{figure}[h!]
    \centering
    \includegraphics[width=\columnwidth]{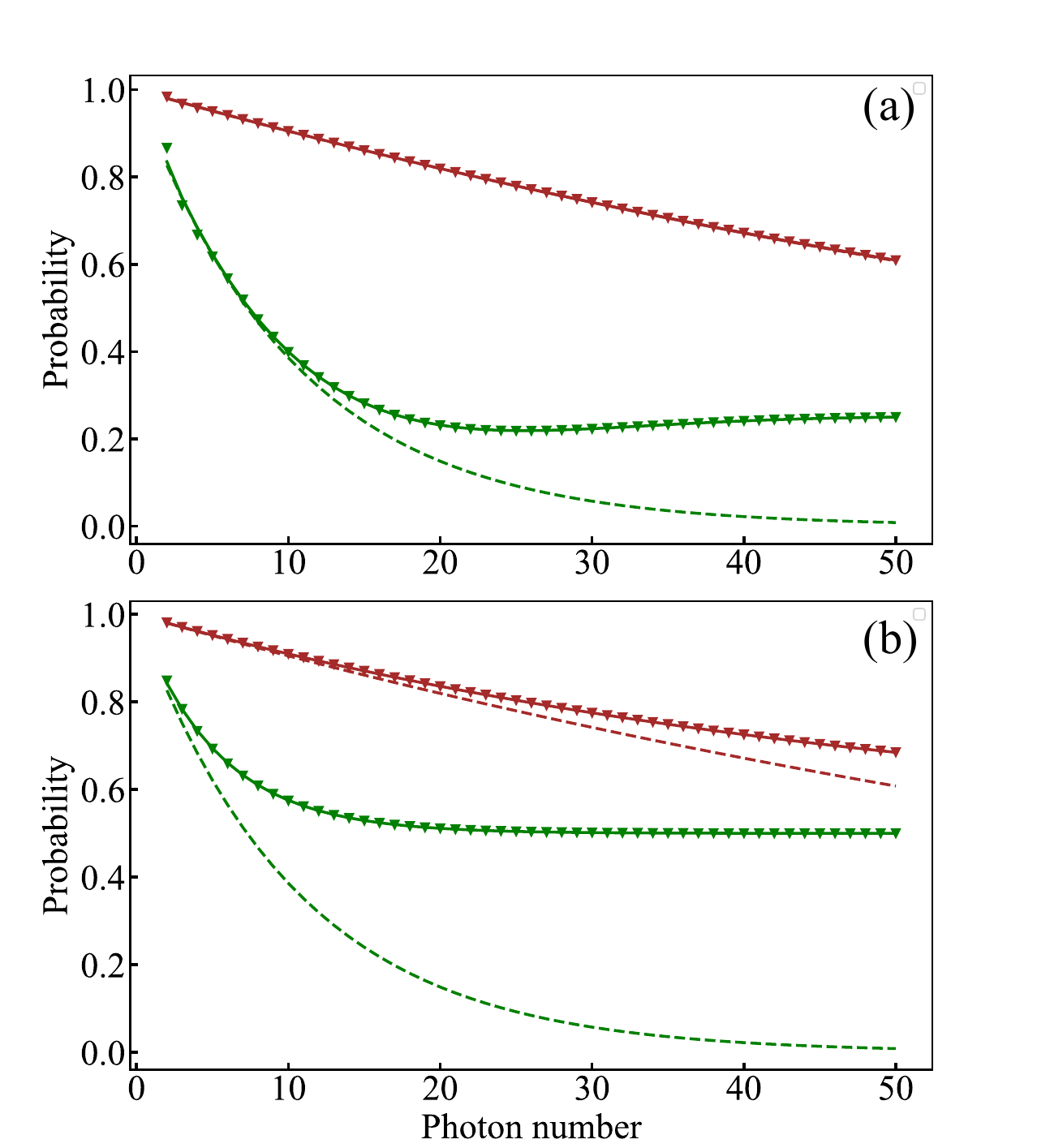}
    \caption{Projection probability (a) $p_0$
    for the logical zero states
    and (b) $p_T$ for the magic states
    against the photon number for $M=2$ in the case of the binomial code. Brown and green indicate the low-error and high-error regimes $\gamma t=0.01$ and $\gamma t=0.1$, respectively. Lined curves, triangles and dashed curves correspond to the exact numerical solution, approximations by (a) $p_0(t) \sim e^{-\Gamma (t)} f_{2M} (\Gamma (t))$ and (b) $p_T(t) \sim e^{-\Gamma (t)} f_{M} (\Gamma (t))$, and approximations by $e^{-\Gamma (t)}$, respectively.}
    \label{Figprojectionbinom}
\end{figure}

We also benchmark the performance of our protocol for the binomial code. Here, we plot the trace distance between the error-free state and the noisy/error-mitigated states for $\gamma t=0.01$ against the photon number of the initial state in Fig. \ref{Figtracedistbinom} for the logical zero and magic states for $M=2$. We can clearly see that the error-mitigated state is closer to the error-free one than the noisy state, which confirms the performance of our protocol. We also numerically calculate the projection probability of SE for the logical zero states and the magic state in Fig. \ref{Figprojectionbinom}. The behavior of the projection probability is akin to the one for the cat code, especially in the large photon regime, which may be attributed to the fact that the photon number distribution of the binomial code comes closer to Poisson distribution in the large photon number limit and these codes asymptotically approaches each other~\cite{michael2016new,albert2018performance}. 

\subsection{Sampling cost for quantum error mitigation} 

From the above discussions, in terms of both the cat code and binomial code, considering the low error regime, when $\bar{n}(1-e^{-\gamma t})\sim \bar{n} \gamma t \lesssim 1 $ for the average photon number $\bar{n}$, the projection probability can be well-approximated as $p \sim e^{-\bar{n} \gamma t}$. Denoting the number of error-mitigated initial states as $N_{\mathrm{QEM}}$, we can see that the variance of the error-mitigated observables is amplified with $C= e^{2 \bar{n} \gamma t N_{\mathrm{QEM}}}$ for the error mitigation of initial states from Eq. (\ref{Eq: amplify}). Therefore, when we only apply symmetry expansion for the initial state preparation, if $\bar{n} \gamma t N_{\mathrm{QEM}} \leq 2$, we can perform quantum error mitigation with the sampling overhead $C \leq 54.6$.

\textcolor{black}{\subsection{Numerical simulation of phase-error suppression}}

To observe the effect of phase-error suppression by the application of the projector $\tilde{P}_{\rm phase}$, we here consider phase noise described by a Lindblad master equation $\frac{d \hat{\rho}}{dt }=\frac{\gamma}{2}(2 \hat{N} \hat{\rho}  \hat{N}- \hat{N}^2\hat{\rho}-\hat{\rho} \hat{N}^2 )$ with $\hat{N}=\hat{a}^\dag \hat{a}$ 
 for four-legged cat states. Note that application of rotation symmetry-based projectors, e.g., $\hat{\mathcal{P}}_{2 M}^{(0)}$ in Eq. (\ref{Eq: projectorrotation}), cannot suppress this type of error because of the commutation relationship $[\hat{N}, \hat{\mathcal{P}}_{2 M}^{(0)}]=0$.  We consider the noisy state under the phase noise for the total evolution time $\gamma t=0.1$. We set $\Gamma=1.5$, which translates into the increased squeezing level $\Delta r= 0.589$. 

As illustrated in the Wigner function in Fig. \ref{Fig:largefig1} (e), the uncertainty of the phase becomes large due to the effect of the phase noise. By applying the smeared projector $\tilde{P}_{\rm phase}$, we confirm that the phase is stabilized.

We then investigate the qubit state constructed from the expectation values of the logical Pauli operators, i.e.,
\begin{equation}
\hat{\rho}^{(L)}=\frac{1}{2}(\hat{I}+r^{(L)}_X \hat{X} + r^{(L)}_Y \hat{Y}+ r^{(L)}_Z \hat{Z}), 
\label{Eq: constructed}
\end{equation}
where $r^{(L)}_X= \mathrm{Tr}[\hat{X}_L \hat{\rho}_{\rm in} ]$, $r^{(L)}_Y= \mathrm{Tr}[\hat{Y}_L \hat{\rho}_{\rm in} ]$ and $r^{(L)}_Z= \mathrm{Tr}[\hat{Z}_L \hat{\rho}_{\rm in} ]$ for the input state $\hat{\rho}_{\rm in}$ with $\hat{X}_L=\sum_{m=0} \ket{m}\bra{m+M}$, $\hat{Z}_L=\mathrm{exp}(i \frac{\pi}{M} \hat{N})$ and $\hat{Y}_L= i \hat{X}_L \hat{Z}_L$. We consider the noisy and the error-mitigated state as the input state $\hat{\rho}_{\rm in}$ with the initial state being the magic state.

\begin{figure}[h!]
    \centering
    \includegraphics[width=1.0\columnwidth]{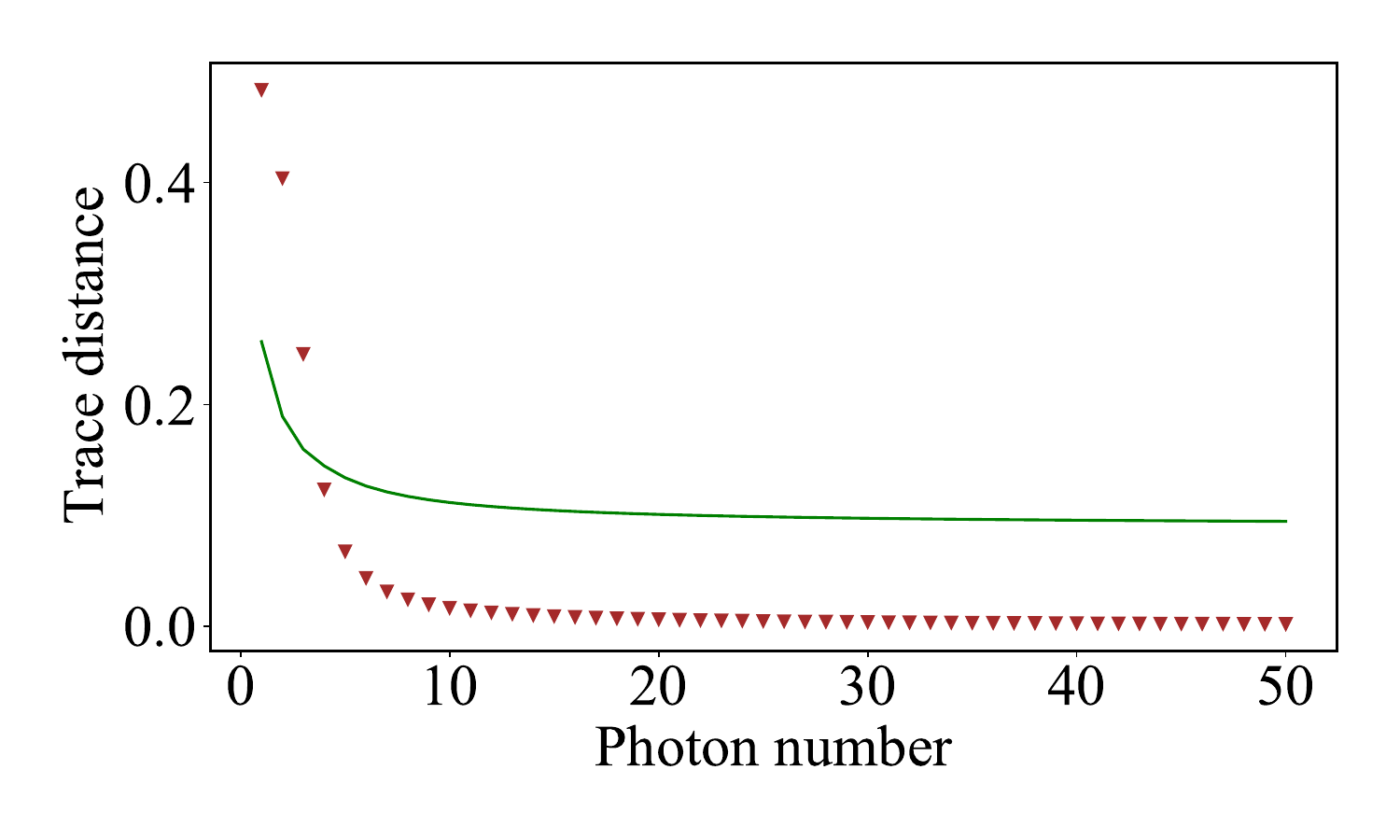}
    \caption{The trace distance between the ideal logical state and the noisy/error-mitigated state under phase noise depending on the photon number $|\alpha|^2$. The green line denotes the noisy results under the phase error, and brown triangles denote the error-mitigated results. }
    \label{Figprojectionps1}
\end{figure}

We plot in Fig. \ref{Figprojectionps1} the trace distance depending on the number of photons $|\alpha|^2$ between the ideal state and the error mitigated state for the constructed qubit state in Eq. \eqref{Eq: constructed}. It is confirmed that the phase error is suppressed because the trace distance is significantly reduced for a large photon number state.

\begin{figure}[h!]
    \centering
    \includegraphics[width=1.0\columnwidth]{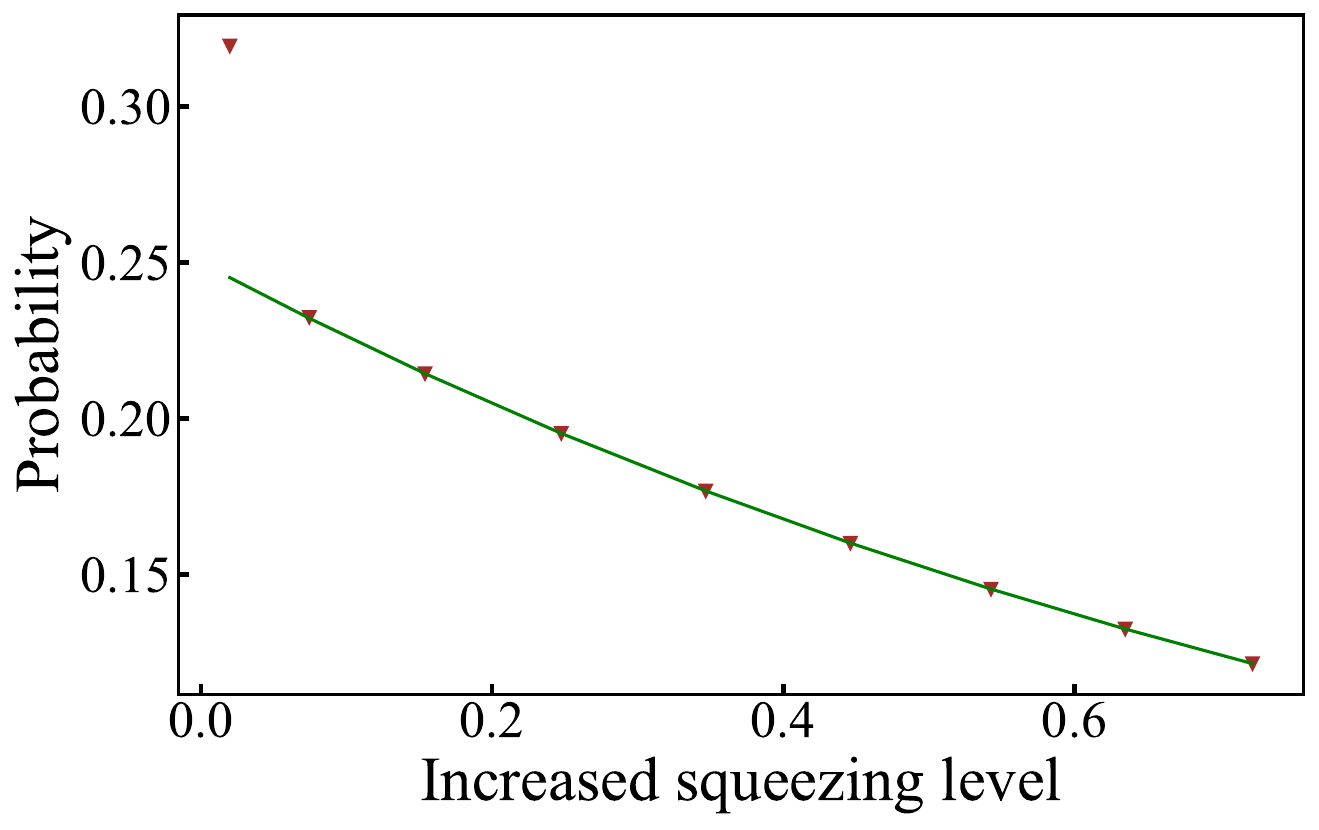}
    \caption{The projection probability for the smeared projector for phase noise-suppression for a four-legged cat code state. The brown triangles denote the numerically calculated projection probability and the green line denote the analytical result in Eq. \eqref{Eq: projectionphase}. }
    \label{Figprojectionps2}
\end{figure}

Furthermore, we plot the numerically calculated projection probability for the input of a noiseless four-legged cat code state. Interestingly, although Eq. (\ref{Eq: projectionphase}) is derived for the superposition of squeezed vacuum states towards orthogonal directions, Eq. (\ref{Eq: projectionphase}) agrees well with the projection probability for a cat code state for a sufficiently large increased squeezing level, as demonstrated in Fig. \ref{Figprojectionps2}.

\section{Conclusion and discussion}
We propose a practical error mitigation technique for RSBCs based on SE, which is especially useful for state preparation. Because state preparation is an important building block for quantum computation using RSBCs due to the necessity for gate operations, our protocol may significantly enhance the overall computation accuracy. For applying SE for state preparation, we introduce a novel method using the generalized quantum process for virtually projecting the quantum state onto the symmetric subspace. For the implementation of SE, two controlled rotation operations are performed by dispersive interactions between an ancilla qubit and the bosonic qubit, followed by the subsequent measurement of the ancilla qubit. Our protocol can circumvent an experimentally demanding verification procedure depending on postselection; thus, it will lead to dramatic noise reduction. With the same formalism for error mitigation, we can virtually generate the RSBC states that give the same expectation value as the ideal one from a primitive state with a sample cost proportional to $M^2$ for the rotation order $M$.

We also benchmark our protocol for practical RSBCs: cat codes and binomial codes. In regard to cat codes, we provide a detailed analytical study
of the projection probability onto the symmetric subspace and the trace distance between the error-mitigated state and the ideal state for photon loss noise. We confirm that the analytical results
coincide well with the exact numerical
solutions. The results of trace distances indicate that the error-mitigated states are significantly closer to the ideal one. In addition, the error mitigation performance improves as $M$ increases. We also evaluate the performance of SE for binomial codes and verify that the noise effects are similarly suppressed. The projection probability for binomial codes approaches the behavior of that of cat codes in the large photon number limit because the photon number distribution becomes approximated by the Poisson distribution. Based on these arguments, we evaluate the sampling overhead of SE in the low error regime, which exponentially increases with the number of error-mitigated qubits $N_{\mathrm{QEM}}$, the effective error rate $\gamma_t= \gamma t$, and the average photon number $\bar{n}$. \textcolor{black}{Note that the error mitigation incurs an exponential increase of the sampling overhead~\cite{takagi2022fundamental,takagi2023universal,tsubouchi2023universal,quek2024exponentially}, and this also holds in our case. Nevertheless, our method is useful in the regime $\bar{n} \gamma t N_{\mathrm{QEM}} =O(1)$, and may find a good regime depending on the development of bosonic quantum devices.}

Finally, the combination of quantum error codes with quantum error mitigation is an active research area~\cite{suzuki2022quantum,xiong2020sampling,piveteau2021error,lostaglio2021error}. An example of other methods than SE is probabilistic error cancellation (PEC)~\cite{temme2017error,endo2018practical} on the code space to effectively increase the code distances~\cite{suzuki2022quantum,xiong2020sampling}, compensate for gate decomposition errors such as approximation errors in Solovay-Kitaev algorithms~\cite{suzuki2022quantum} and  T gate errors~\cite{suzuki2022quantum,piveteau2021error,lostaglio2021error}. Investigating the relationship between SE and PEC in the code space is an interesting direction. For example, seeing whether PEC can be combined with SE for further improvement of computation accuracy for bosonic code quantum computation could be practically important.

\vspace{5mm}

\section*{Acknowledgments}
This project is supported by Moonshot R\&D, JST, Grant No.\,JPMJMS2061; MEXT Q-LEAP Grant No.\,JPMXS0120319794, and No.\, JPMXS0118068682 and PRESTO, JST, Grant No.\, JPMJPR2114, No. \,
JPMJPR1919 and  No. \,JPMJPR1916.
This work was supported by Leading Initiative for Excellent Young Researchers MEXT Japan.
S.E thanks useful discussions with Hideaki Hakoshima. Nobuyuki Yoshioka, and Zhenyu Cai. Numerical simulations were performed with QuTip~\cite{johansson2012qutip}. K.T. is supported by the Program for Leading Graduate Schools (MERIT-WINGS) and JST BOOST Grant Number JPMJBS2418.

\bibliography{bib}

\newpage

\appendix

\onecolumngrid

\section{Error-mitigated state and the projection probability}
\label{Appendix: A}
To clarify the performance of our protocol, it is important to evaluate the error-mitigated state and projection probabilities for important states such as $\ket{0_{M, \Phi}}$, $\ket{T_{M, \Phi}}$. The Lindblad master equation for photon loss $\frac{d}{dt} \rho(t)=\frac{\gamma}{2} (2 \hat{a} \hat{\rho} \hat{a}^\dag -\hat{a}^\dag \hat{a} \hat{\rho}- \hat{\rho} \hat{a}^\dag \hat{a})$ leads to \cite{bachor1992atomic,cochrane1999macroscopically,chuang1997bosonic}:
\begin{equation}
\begin{aligned}
\hat{\rho}(t)&=\sum_{L=0}^\infty \hat{\rho}_L(t) \\
\hat{\rho}_L(t)& = \frac{(1-e^{-\gamma t})^L}{L!} e^{-(\gamma t/2) \hat{N} } \hat{a}^L \hat{\rho}(0)  (\hat{a}^{\dag})^L e^{-(\gamma t/2) \hat{N} }.
\label{Eq: photondamp}
\end{aligned}
\end{equation}
When we set the initial state to a superposition state of coherent states $\ket{\psi(0)}=\sum_m c_m \ket{\alpha_m}$, we can easily obtain
\begin{equation}
\hat{\rho}(t)=\sum_{m m'} c_m c_{m'}^{*} \mathrm{exp}((1-e^{-\gamma t})\alpha_m\alpha_{m'}^* )\mathrm{exp}\bigg(-\frac{|\alpha_m|^2+|\alpha_{m'}|^2}{2} (1-e^{-\gamma t}) \bigg) \ket{\alpha_m e^{-\gamma t/2}}  \bra{\alpha_{m'} e^{-\gamma t/2}}.
\end{equation}
\subsection{Error mitigation for $\ket{0_{M, \alpha}}$}
Considering the case of $\ket{\psi(0)}=\ket{0_{M, \alpha}}$ for $\ket{\Phi}=\ket{\alpha}$, we have

\begin{equation}
\hat{\rho}(t) = \frac{1}{\mathcal{N}}\sum_{m, m'=0}^{2M-1} \mathrm{exp}(-\Gamma (t)) e^{\Gamma (t) e^{i\pi\frac{m-m'}{M}}} \ket{\alpha (t) e^{i\frac{m}{M} \pi}  } \bra{\alpha (t) e^{i\frac{m'}{M} \pi}  }.
\end{equation}

Here, $\mathcal{N}$ is the time-independent normalization factor, $\alpha(t)=\alpha e^{-\gamma t/2}$ and $\Gamma (t)= \alpha^2 (1-e^{-\gamma t})$. Since $\hat{\mathcal{P}}_{2M}^0 \ket{\alpha (t) e^{i\frac{m}{M} \pi} } \propto \ket{0_{M, \alpha(t)}} \forall m$, we can clearly see 
\begin{equation}
\hat{\mathcal{P}}_{2 M}^{(0)} \hat{\rho}(t) \hat{\mathcal{P}}_{2 M}^{(0)} \propto \ket{0_{M, \alpha (t)}}\bra{0_{M, \alpha (t)}},
\end{equation}
which indicates that SE projects the noisy state onto the symmetric subspace of the RSBCs. 

Now, we discuss the projection probability $p_0$. We first calculate the normalization factor as 
\begin{equation}
\begin{aligned}
\mathcal{N} &= \sum_{m, m'=0}^{2M-1} \braket{\alpha e^{i\frac{\pi m'}{M}}| \alpha e^{i\frac{\pi m}{M}}  } \\
&=e^{-\alpha^2} \sum_{m, m'=0}^{2M-1} e^{\alpha^2 \mathrm{exp} (-i \frac{\pi (m'-m)}{M})} \\
&= 2M e^{-\alpha^2} \sum_{m=0}^{2M-1} e^{\alpha^2 \mathrm{exp}(i\frac{\pi m}{M})}.
\end{aligned}
\end{equation}

Here, in the second line we used $\braket{\beta|\beta'}= \mathrm{exp}(-\frac{1}{2} |\beta|^2-\frac{1}{2}|\beta'|^2+\beta^* \beta')$ for two coherent states $\ket{\beta}$ and $\ket{\beta'}$. In the third line, we used the periodicity of the exponential function. Then, with Taylor's expansion, we have
\begin{equation}
\begin{aligned}
\sum_{m=0}^{2M-1} e^{\alpha^2 \mathrm{exp}(i\frac{\pi m}{M})}&= \sum_{m=0}^{2M-1} \sum_{l=0} \frac{1}{l!} \alpha^{2 l} \mathrm{exp}\big(i \frac{\pi m l}{M} \big) \\
&=2M \sum_{l=0} \frac{(\alpha^2)^{2Ml}}{(2Ml)!} =2M f_{2M}(\alpha^2),
\label{Eq: fmfunc}
\end{aligned}
\end{equation}
where we define $f_M(x):= \sum_{l=0} \frac{(x)^{Ml}}{(Ml)!}$ and employ $\sum_{m=0}^{2M-1} \mathrm{exp}(i\frac{\pi l m}{M})=2M$ for $l=2M l' ~(l'=0,1,2,...)$, otherwise $0$. Thus, we have
\begin{equation}
\mathcal{N}= 4M^2 e^{-\alpha^2} f_{2M} (\alpha^2),
\end{equation}
which converges to $2M$ for sufficiently large $\alpha^2$. Since we have $\mathrm{lim}_{\alpha \rightarrow \infty} \mathcal{N}=2M$
, we get $\mathrm{lim}_{\alpha \rightarrow \infty} f_{2M}(\alpha^2)=e^{\alpha^2}/2M$.
The projection probability $p_0$ can be written as 
\begin{equation}
\begin{aligned}
p_0&= \mathrm{Tr}[\hat{\mathcal{P}}_{2M}^{(0)} \hat{\rho}(t) \hat{\mathcal{P}}_{2M}^{(0)}] \\
&= \frac{1}{\mathcal{N}}\sum_{m, m'=0}^{2M-1} \mathrm{exp}(-\Gamma (t)) e^{\Gamma (t) e^{i\frac{m-m'}{M}}\pi} \bigg(\frac{1}{2M}\bigg)^2 \\
&\times \sum_{k, k'=0}^{2M-1} \braket{\alpha (t) e^{i\frac{k'}{M} \pi}|\alpha (t) e^{i\frac{k}{M} \pi} }. 
\end{aligned}
\end{equation}
 A similar argument for the calculation of the normalization factor leads to 
 \begin{equation}
\sum_{m, m'=0}^{2M-1}  e^{\Gamma (t) e^{i\frac{m-m'}{M}}\pi} = 4M^2  f_{2M}(\Gamma (t)),
 \end{equation}
and 
\begin{equation}
\mathcal{N}':=\sum_{k, k'=0}^{2M-1} \braket{\alpha (t) e^{i\frac{k'}{M} \pi} |\alpha (t) e^{i\frac{k}{M} \pi} }=4M^2 e^{-\alpha(t)^2} f_{2M}(\alpha (t)^2 ).
\end{equation}
Note that $\mathcal{N}'$ converges to $2M$ when $\alpha (t)^2$ is sufficiently large.  Therefore, in the regime where the photon count is large enough and the error rate is low, we obtain the approximated projection probability:
\begin{equation}
p_0 \sim e^{-\Gamma (t)} f_{2M} (\Gamma (t)).
\end{equation}
The projection probability $p_0$ can be further approximated as $p_0 \sim e^{-\Gamma (t)}\sim e^{-\alpha^2 \gamma t}$ under the assumption of $f_{2M}(\Gamma (t)) \sim 1$ for $\alpha^2 \gamma t \ll 1$. Meanwhile, the exact analytical expression of the projection probability can be obtained as follows:
 \begin{equation}
p_0=\frac{f_{2M}(\Gamma (t)) f_{2M}(\alpha (t)^2 )}{f_{2M}(\alpha^2)}. 
 \end{equation}
 
\subsection{Error mitigation for general logical qubit states}
 Here, we derive the error mitigated state and the projection probability for the general qubit state $\ket{\psi_{M,\Phi}}=a \ket{0_{M,\Phi}}+b \ket{1_{M,\Phi}}$ under photon loss. The density matrix for the noiseless state can be described as
 \begin{equation}
\hat{\rho}_\psi (0)=\sum_{m, m'=0}^{2M-1} c_m c_{m'}^* \ket{\alpha e^{i\frac{\pi m}{M}}}\bra{\alpha e^{i\frac{\pi m'}{M}}},
 \end{equation}
 where $c_{m}=\frac{a}{\sqrt{\mathcal{N}_0}}+(-1)^m \frac{b}{\sqrt{\mathcal{N}_1}}$ with $\mathcal{N}_0$ and  $\mathcal{N}_1$ being the normalization factors for $\ket{0_{M, \Phi}}$ and $\ket{1_{M, \Phi}}$. Hereafter, we assume $\mathcal{N}_0=\mathcal{N}_1=2M$ by assuming a sufficiently large photon count. Under the photon loss noise described by Eq. \eqref{Eq: photondamp}, we have the state at time $t$:
 \begin{equation}
\hat{\rho}_\psi(t)=\sum_{m, m'=0}^{2M-1} c_m c_{m'}^* e^{-\Gamma (t)} e^{\Gamma (t)e^{i\pi\frac{m-m'}{M}}}   \ket{\alpha (t) e^{i\frac{m}{M} \pi}  } \bra{\alpha (t) e^{i\frac{m'}{M} \pi} }.
 \end{equation}
  Then we apply the projector onto the code space $\hat{\mathcal{P}}_{2M}^{(\mathrm{c})}=\frac{1}{M} \sum_{k=0}^{M-1} e^{i \frac{2\pi k}{M} \hat{N}}$ to the state $\hat{\rho}_\psi(t)$ to get
 \begin{equation}
 \begin{aligned}
\hat{\mathcal{P}}_{2M}^{(\mathrm{c})} \hat{\rho}_\psi(t) \hat{\mathcal{P}}_{2M}^{(\mathrm{c})} &=\frac{1}{M}(  |c_0|^2 \sum_{k,k'=0}^{M-1} e^{-\Gamma (t)}e^{\Gamma (t) e^{i \frac{2(k-k')}{M} \pi}} \ket{+(t)}\bra{+(t)} \\
&+c_0 c_1^* \sum_{k,k'=0}^{M-1} e^{-\Gamma (t)}e^{\Gamma (t) e^{i \frac{2(k-k')-1}{M} \pi}} \ket{+(t)}\bra{-(t)} \\
&+c_0^* c_1 \sum_{k,k'=0}^{M-1} e^{-\Gamma (t)}e^{\Gamma (t) e^{i \frac{2(k-k')+1}{M} \pi}} \ket{-(t)}\bra{+(t)} \\
&+|c_1|^2 \sum_{k,k'=0}^{M-1} e^{-\Gamma (t)}e^{\Gamma (t) e^{i \frac{2(k-k')}{M} \pi}} \ket{-(t)}\bra{-(t)}), \\
\end{aligned}
\end{equation}
where $\ket{+(t)}=  \frac{1}{\sqrt{M}} \sum_{k=0}^{M-1} e^{i \frac{2\pi k}{M} \hat{N}} \ket{\alpha (t) } \sim \frac{1}{\sqrt{2}} (\ket{0_{M, \alpha (t) }}+ \ket{1_{M, \alpha (t) }})$ and  $\ket{-(t)}=  \frac{1}{\sqrt{M}} \sum_{k=0}^{M-1} e^{i \frac{2\pi k}{M} \hat{N}} \ket{e^{i\frac{\pi}{M}} \alpha(t) } \sim \frac{1}{\sqrt{2}} (\ket{0_{M, \alpha(t) }}- \ket{1_{M, \alpha (t) }})$. Then, we obtain 
\begin{equation}
\begin{aligned}
\hat{\mathcal{P}}_{2M}^{(\mathrm{c})} \hat{\rho}_\psi(t) \hat{\mathcal{P}}_{2M}^{(\mathrm{c})}&= Me^{-\Gamma (t)}(|c_0|^2 f_M(\Gamma (t))\ket{+(t)}\bra{+(t)}+c_0 c_1^* g_M(\Gamma (t))\ket{+(t)}\bra{-(t)} \\
&+c_0^* c_1 g_M(\Gamma (t)) \ket{-(t)}\bra{+(t)}+|c_1|^2 f_M(\Gamma (t))\ket{-(t)}\bra{-(t)} ),
\end{aligned}
\end{equation}
 where $g_M(x)= \sum_{l=0} \frac{x^{Ml}}{Ml!}(-1)^l$. 
Here, we use
\begin{equation}
\begin{aligned}
\frac{1}{M}\sum_{k,k'=0}^{M-1} e^{\Gamma(t) \mathrm{exp}(i\frac{2 (k-k')\pi}{M})}
&=\sum_{m=0}^{M-1} e^{\Gamma(t) \mathrm{exp}(i\frac{2\pi m}{M})}\\
&= \sum_{m=0}^{M-1} \sum_{l=0} \frac{1}{l!} (\Gamma (t))^{l} \mathrm{exp}\big(i \frac{2\pi m l}{M} \big) \\
&=M \sum_{l=0} \frac{(
\Gamma(t)
)^{Ml}}{(Ml)!} =M f_{M}(
\Gamma(t)
),
\label{Eq: fmfunc}
\end{aligned}
\end{equation}
and
\begin{equation}
\begin{aligned}
\frac{1}{M}\sum_{k,k'=0}^{M-1} e^{\Gamma(t) \mathrm{exp}(i\frac{2 (k-k')\pm1}{M}\pi)}
&=\sum_{m=0}^{M-1} e^{\Gamma(t) \mathrm{exp}(i\frac{2 m \pm 1}{M}\pi)}\\
&= \sum_{m=0}^{M-1} \sum_{l=0} \frac{1}{l!} (\Gamma(t))^l \mathrm{exp}\big(i \frac{(2m\pm 1)\pi  l}{M} \big) \\
&=M \sum_{l=0} \frac{(\Gamma(t))^{Ml}}{(Ml)!}
(-1)^l
=M g_{M}(\Gamma(t)  
).
\label{Eq: fmfunc2}
\end{aligned}
\end{equation}

 Then, with proper normalization, we get the error-mitigated state:
 \begin{equation}
\begin{aligned}
\hat{\rho}_\psi^{\rm EM}(t)&= \frac{|c_0|^2}{|c_0|^2+|c_1|^2}\ket{+(t)}\bra{+(t)}+\frac{c_0 c_1^*}{|c_0|^2+|c_1|^2}\frac{g_M(\Gamma (t) )}{f_M(\Gamma (t) )}\ket{+(t)}\bra{-(t)} \\
&+\frac{c_0^* c_1}{|c_0|^2+|c_1|^2}\frac{g_M(\Gamma (t) )}{f_M(\Gamma (t) )}\ket{-(t)}\bra{+(t)}+\frac{|c_1|^2}{|c_0|^2+|c_1|^2}\ket{-(t)}\bra{-(t)}.
\end{aligned}
 \end{equation}
 For the magic state $\ket{T_{M,\Phi}}=\frac{1}{\sqrt{2}}(\ket{0_{M,\Phi}}+e^{i\pi/4}\ket{1_{M,\Phi}})$, we have
 \begin{equation}
\begin{aligned}
\hat{\rho}_T^{\rm EM}(t)&= \frac{1}{2}\bigg(1+\frac{1}{\sqrt{2}}\bigg)\ket{+(t)}\bra{+(t)}+i\frac{1}{2\sqrt{2}}\frac{g_M(\Gamma (t) )}{f_M(\Gamma (t) )}\ket{+(t)}\bra{-(t)} \\
&-i\frac{1}{2\sqrt{2}}\frac{g_M(\Gamma (t) )}{f_M(\Gamma (t) )}\ket{-(t)}\bra{+(t)}+\frac{1}{2}\bigg(1-\frac{1}{\sqrt{2}}\bigg)\ket{-(t)}\bra{-(t)}.
\end{aligned}
 \end{equation}
 
For the state $\ket{+i_{M,\Phi}}=\frac{1}{\sqrt{2}}(\ket{0_{M,\Phi}}+i\ket{1_{M,\Phi}})$, we get

\begin{equation}
\begin{aligned}
\hat{\rho}_{+i}^{\rm EM}(t)&= \frac{1}{2}\ket{+(t)}\bra{+(t)}+\frac{i}{2}\frac{g_M(\Gamma (t) )}{f_M(\Gamma (t) )}\ket{+(t)}\bra{-(t)} \\
&-\frac{i}{2}\frac{g_M(\Gamma (t) )}{f_M(\Gamma (t) )}\ket{-(t)}\bra{+(t)}+\frac{1}{2}\ket{-(t)}\bra{-(t)}.
\end{aligned}
 \end{equation}
 
The error-mitigated state for the plus state $\ket{+_{M,\Phi}}=\frac{1}{\sqrt{2}}(\ket{0_{M,\Phi}}+\ket{1_{M,\Phi}})$ becomes 
\begin{equation}
\hat{\rho}_{+}^{\rm EM}(t)= \ket{+(t)}\bra{+(t)}.
\end{equation}
This is because we have $\ket{0_{M, \alpha}} \sim \ket{+_{2M, \alpha}} $ in the large photon number limit.

 On the other hand, we get the projection probability 
 \begin{equation}
\begin{aligned}
 p_\psi&=\mathrm{Tr}[\hat{\mathcal{P}}_{2M}^{(\mathrm{c})} \hat{\rho}_\psi(t) \hat{\mathcal{P}}_{2M}^{(\mathrm{c})}] \sim M(|c_0|^2+|c_1|^2)e^{-\Gamma(t)}  f_M(\Gamma (t))  \\
 &=e^{-\Gamma (t)} f_M(\Gamma (t)),
\end{aligned}
 \end{equation}
where we assume $\braket{+(t)|-(t)}\sim 0$. For $\alpha^2 \gamma t \ll 1$,  the projection probability $p_\psi$ can be further approximated as $p_\psi \sim e^{-\Gamma (t)} \sim e^{-\alpha^2 \gamma t}$.

\section{The regime where the coherent states are distinguishable}
\label{Appendix: B}
The two coherent states $\ket{\alpha}$ and $\ket{-\alpha}$  ($\alpha \in \mathbb{R}$) are sufficiently distinguishable for $|\alpha|^2 \gtrsim 1.5$ with the overlap of two states being approximately $0.2 \%$. We denote such an amplitude by $\alpha_0$. Then, we consider the regime where coherent states constituting the RSCB with the rotation degree $M$ have a low enough overlap. The overlap of the $\ket{\alpha_M}$ and $\ket{\alpha_M e^{i\pi/M}}$ ($\alpha_M \in \mathbb{R}$) should be
\begin{equation}
|\braket{\alpha_M|\alpha_M e^{i\pi/M}}|^2 \lesssim |\braket{\alpha_0 | -\alpha_0}|^2,
\end{equation}
which leads to
\begin{equation}
|\alpha_M|^2 \gtrsim \frac{|\alpha_0|^2}{\mathrm{sin}^2\big(\frac{\pi}{2M}\big)}.
\end{equation}
For example, we have $|\alpha_{M=2}|^2\gtrsim 3$ and 
$|\alpha_{M=4}|^2 \gtrsim 10.2$, 
which are used in the plot in Fig. \ref{Figtracedist}.

\section{Phase error suppression via phase stabilizers}

\label{sec:supplec}

\begin{figure}[h!]
    \centering
    \includegraphics[width=0.5\columnwidth]{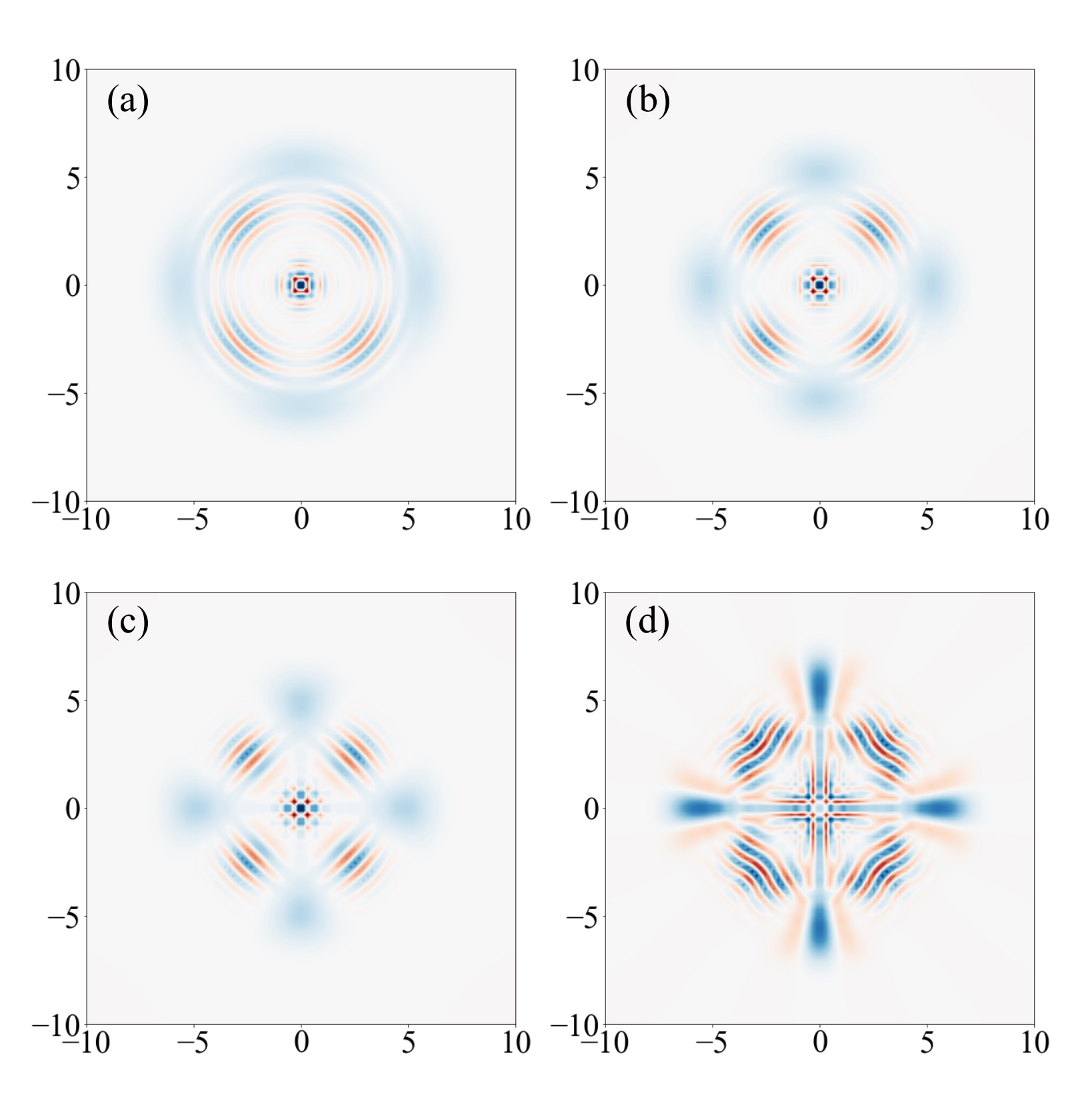}
    \caption{Wigner functions of noisy cat states under phase noise (a) and error-mitigated states via a truncated projector $\hat{\mathcal{P}}_X^{(L)}$ (b)-(d). Figs (b), (c), and (d) show the cases of $L=1$, $L=2$, and $L=3$, respectively. We set $\gamma t =0.1$. 
    }
    \label{Fig:phasecorrect}
\end{figure}

To clarify the complementary effect of $\hat{\mathcal{P}}_X^{(L)}$ compared with the projector obtained from rotation symmetries, we here consider phase noise described by a Lindblad master equation $\frac{d \hat{\rho}}{dt }=\frac{\gamma}{2}(2 \hat{N} \hat{\rho}  \hat{N}- \hat{N}^2\hat{\rho}-\hat{\rho} \hat{N}^2 )$. Note that application of rotation symmetry-based projectors, e.g., $\hat{\mathcal{P}}_{2 M}^{(0)}$ in Eq. (\ref{Eq: projectorrotation}), cannot suppress this type of error because of the commutation relationship $[\hat{N}, \hat{\mathcal{P}}_{2 M}^{(0)}]=0$. Meanwhile, we found truncated projectors can mitigate phase noise, as shown in Fig. \ref{Fig:phasecorrect}. The error-mitigation effect can be enhanced as we increase the truncation level $L$. This is because the projector $\hat{\mathcal{P}}_X^{(L)}$ has the effect of projecting the state onto the subspace in which the phase is fixed, as indicated by the fact that phase errors can be detected through the logical $X$ measurement. 

\end{document}